\documentclass[twocolumn,tighten]{aastex62}

\usepackage{graphicx}
\usepackage{bm}
\usepackage{url}
\usepackage{threeparttable}
\usepackage{amsmath,amssymb}
\usepackage{xspace}
\usepackage{ulem}
\usepackage{subfigure}

\newcommand{\spitzer}{\textit{Spitzer}\xspace}
\newcommand{\rosat}{\textit{ROSAT}\xspace}
\newcommand{\molline}[3]{#1($J$=#2--#3)}
\newcommand{\chandra}{{\it Chandra}\xspace}

\newcommand{\ergcms}{erg cm$^{-2}$ s$^{-1}$}
\newcommand{\ergs}{erg s$^{-1}$} 
\newcommand{\modtext}[1]{{\textcolor{black}{#1}}} 

\shorttitle{X-ray irradiation of CO gas in NGC~2110} 
\shortauthors{Kawamuro et al.}

\begin{document}
\title{
  AGN X-ray irradiation of CO gas in NGC~2110 revealed by \chandra and ALMA
} 

\correspondingauthor{Taiki Kawamuro}
\email{taiki.kawamuro@nao.ac.jp}

\author{Taiki Kawamuro}
\altaffiliation{JSPS fellow (PD)}
\affil{National Astronomical Observatory of Japan, Osawa, Mitaka, Tokyo 181-8588, Japan}

\author{Takuma Izumi}
\altaffiliation{NAOJ fellow} 
\affil{National Astronomical Observatory of Japan, Osawa, Mitaka, Tokyo 181-8588, Japan}
\affil{Department of Astronomy, School of Science, Graduate University for Advanced Studies (SOKENDAI), 2-21-1 Osawa, Mitaka, Tokyo 181-8588}

\author{Kyoko Onishi} 
\affil{Department of Earth and Space Sciences, Chalmers University of Technology, Onsala Observatory, 439 94 Onsala, Sweden}

\author{Masatoshi Imanishi} 
\affil{National Astronomical Observatory of Japan, Osawa, Mitaka, Tokyo 181-8588, Japan}
\affil{Department of Astronomy, School of Science, Graduate University for Advanced Studies (SOKENDAI), 2-21-1 Osawa, Mitaka, Tokyo 181-8588}

\author{Dieu D. Nguyen}
\affil{National Astronomical Observatory of Japan, Osawa, Mitaka, Tokyo 181-8588, Japan}

\author{Shunsuke Baba}
\altaffiliation{JSPS fellow (PD)}
\affil{National Astronomical Observatory of Japan, Osawa, Mitaka, Tokyo 181-8588, Japan}



\begin{abstract}

  We report spatial distributions of the Fe-K$\alpha$ line at 6.4 keV and the CO($J$ = 2--1) line at 230.538~GHz in NGC~2110, which are respectively revealed by \chandra and ALMA at $\approx$ 0.5 arcsec. A \chandra 6.2--6.5~keV-to-3.0--6.0~keV image suggests that the Fe-K$\alpha$ emission extends preferentially in a northwest-to-southeast direction out to $\sim$ 3 arcsec, or 500 pc, on each side. Spatially-resolved spectral analyses support this by finding significant Fe-K$\alpha$ emission lines only in northwest and southeast regions. Moreover, their equivalent widths are found $\sim$ 1.5~keV, indicative for the fluorescence by nuclear X-ray irradiation as the physical origin. By contrast, CO($J$ = 2--1) emission is weak therein. For quantitative discussion, we derive ionization parameters by following an X-ray dominated region (XDR) model. We then find them high enough to interpret the weakness as the result of X-ray dissociation of CO and/or H$_2$. Another possibility also remains that CO molecules follow a super-thermal distribution, resulting in brighter emission in higher-$J$ lines. Further follow-up observations are encouraged to draw a conclusion on what predominantly changes the inter-stellar matter properties, and whether the X-ray irradiation eventually affects the surrounding star formation as an AGN feedback.
  
  
  
  
\end{abstract}


\section{Introduction}

Harsh radiation from a mass accreting super-massive black hole (SMBH), or an active galactic nucleus (AGN), can change thermal and chemical properties of the surrounding inter-stellar medium (ISM). The SMBH is  usually present in the center of a massive galaxy, and it has been suggested that the SMBH and host galaxy have grown while affecting each other  \citep[i.e., the co-evolution; e.g., ][]{Mag98,Mar03,Geb00,Fer00,Kor13}. This implies that the host galaxy has formed stars while being subject to the AGN radiation. Thus, towards the complete understanding of the galaxy growth, the study of the AGN radiative effect is important.  


The AGN is more X-ray luminous than stars, and affects the ISM in a  fundamentally different way. Also, because of the high energy of X-ray photons, the AGN X-ray emission is expected to largely affect ISM properties. Such a region subject to the X-ray emission is conventionally referred to as the X-ray dominated region \citep[XDR; e.g.,  ][]{Kro83,Lep96,Mal96,Mal99}, and has been often studied theoretically \citep{Use04,Mei05,Mei07,Pro14}. There was even a prediction that the X-ray emission changes the initial stellar mass function by evaporating a thinner part of a gas cloud and compressing its  thicker part \citep{Hoc10,Hoc11}. Among the predictions compared with  observational works \citep[e.g., ][]{Gal10,Izu15,Kaw19b}, a noticeable point regarding the co-evolution would be the X-ray dissociation of molecular gas. Given a good correlation between the star-forming region and molecular gas distribution \citep[e.g., ][]{Ken07,Big08} and theoretical arguments \citep[e.g., ][]{Glo12,Byr19}, star formation can be active in regions where hydrogen molecules are efficiently produced. Thus, qualitatively, X-ray emission that dissociates molecular gas may work to inhibit star formation, or as a negative AGN feedback.


X-ray-irradiated regions have been often probed with \chandra \citep[e.g., ][]{You01,Wan09,Mar12,Mar13,Mar17,Gom17,Fab17,Fab19b,Fab19a,Kaw19b} by exploiting its high angular resolution ($<$ 0.5 arcsec). A simple way to unveil such regions is to map the Fe-K$\alpha$ fluorescent line at 6.4 keV, originated by the ionization of a K-shell electron due to an X-ray above the 7.1 keV edge energy. In comparison with focusing on soft X-ray emission, adopted in many studies, we can probe  X-ray-irradiated denser gas. This is because 
the hard X-ray emission that produces the Fe emission can penetrate into the gas deeply, and the Fe emission can break out of the gas due to the high penetrating power. Little contamination from stellar light above 7.1 keV \citep[e.g., ][]{LaM12,Kaw13,Lam17} is also an advantage to purely trace regions subject to AGN emission.

In this paper, we discuss ISM properties in the central $\approx$ 1.3 kpc of NGC~2110, which hosts an obscured AGN, based on the Fe-K$\alpha$ and CO($J$ = 2--1) emission lines. Recently, \cite{Ros19} found a region with weak CO($J$ = 2--1) emission within $\sim$ 600 pc of NGC~2110. This study was then followed by \cite{Fab19b}, who reported the presence of soft X-ray photons in the region \citep[see also][]{Eva06}. Following them, we provide further pieces of information obtained by constraining the Fe-K$\alpha$ distribution, eventually enabling quantitative discussion for the weak CO($J$ = 2--1) emission. 

This paper is organized as follows. In Section~\ref{sec:ngc2110}, we briefly describe what were previously found and suggested for NGC~2110. Then,  Sections~\ref{sec:obsdata} and \ref{sec:datana} present a summary of our \chandra and ALMA datasets and their analyses, respectively. Section~\ref{sec:dis} is dedicated to discussion. Finally, the summary of this paper is presented in Section~\ref{sec:sum}. Unless otherwise noted, errors are quoted at the 1$\sigma$ confidence level for a single parameter of interest.



\section{NGC 2110}\label{sec:ngc2110}

NGC~2110 is located at a redshift of 0.00779 ($cz$ = 2335$\pm$20 km s$^{-1}$), determined from an optical Mg $b$ absorption line by \cite{Nel95}. Under the assumption of a $\Lambda$CDM cosmology with $H_0$ = 70 km m$^{-1}$ Mpc$^{-1}$, $\Omega_{\rm m} = 0.3$, and $\Omega_\Lambda = 0.7$, the luminosity and angular distances are 33.6 Mpc and 33.0 Mpc  (i.e., 1 arcsec  $\approx$ 160 pc). An optical center of the galaxy is (R.A., Dec. = 88.047420, $-$7.456212 = 5h52m11.381s, $-$7d27m22.36s) \citep{Cle83}. The galaxy is categorized as an early type S0 galaxy \citep{Dev91}, and the stellar mass was estimated to be $\sim 10^{10.6}~M_{\rm sun}$ \citep{Kos11}. Also, molecular and atomic hydrogen gases distribute on scales of $\sim$ 200--600 pc, and their masses would amount to $\sim 10^8~M_{\rm sun}$ and $\sim 10^7~M_{\rm sun}$, respectively \citep[e.g., ][]{Gal99,Ros19}.

NGC~2110 hosts a type-2 AGN \citep{Bra78,McC79,Shu80}. The detection of polarized broad H$\alpha$ emission by \cite{Mor07} suggests the presence of a type-1-AGN-like core, however obscured by dust. Because of the obscuration, the nuclear structure was often investigated by X-ray observations since \cite{Bra78}. Observed X-ray spectra showed Fe-K$\alpha$ emission with a velocity width  of $\sim$ 2000 km s$^{-1}$ \citep{Shu10} and time-variable absorptions with column densities of $N_{\rm H} \sim {\rm a~few}\times10^{22}$ cm$^{-2}$ \cite[e.g., ][]{Bra78,Hay96,Riv14,Mar15,Kaw16b}. Thus, an  amount of matter is likely located around the nucleus, and may form a putative torus. Absorption-corrected 2--10 keV luminosities were measured to be $\sim 10^{42-43}$ \ergs \cite[e.g., ][]{Mus82,Wea95,Riv14,Kaw18}. 




NGC~2110 has been a good target to study impacts of the jet and nuclear radiation on the ISM simultaneously. Indeed, a radio jet in a north-south direction is clearly seen from parsec to $\approx$ 600 pc scales  \citep[e.g., ][]{Ulv83,Nag99,Mun00}. For example, an impact of the jet on the ISM was indicated by NIR [Fe~II] emission bright in the north-south direction \citep[e.g., ][]{Sto99,Dur14,Din15,Din19}. On the other hand, regions subject to the nuclear UV radiation were suggested from distributions of hydrogen emission lines extending with an angle $\sim 30$ degrees from north to west \citep[e.g., ][]{Wil85,Pog89,Mul94,Del02,Fer04}. Moreover, soft X-ray ($<~2$~keV) emission found out to $\sim$ a few kpc would suggest that the nuclear X-ray emission  also contributes to excitation of ambient gas \citep{Wea95,Eva06,Fab19b}.

\section{Observation data}\label{sec:obsdata}

\subsection{\chandra data}\label{sec:cxodata}

We analyzed one \chandra/ACIS-S (\citealt{Gar03}) imaging data of NGC~2110 (ObsID = 883)\footnote{Although NGC 2110 was observed three times via the High-Energy Transmission Grating (\citealt{Can05}), the observed data were not utilized in this study. This was because the data in the 0th order, equivalent to imaging, seem to have non-negligible inconsistency regarding spatial distributions with simulated data by the MARX software, which played an important role in our analysis (Section~\ref{sec:ironmap}).} to probe X-ray-irradiated regions. Our analysis utilized the standard data analysis package CIAO (ver.~4.9) and a calibration database of CALDB (ver. 4.7.6). The raw data was reprocessed with the standard \texttt{chandra\_repro} command. Periods with slight increase of background rates were found during the observation and were filtered out. An exposure of $\approx$ 45~ksec was left for further analyses. The pile-up effect, where more than one photons are counted as one photon within a single readout, was checked by using \texttt{pileup\_map}. Only in the central 1 arcsec region (Figure~\ref{fig:x_ima}),  pile-up fractions were larger than 5\%. Because our interest is in spatially-extended emission, we do not further discuss the central region.


Accuracy of the absolute coordinates of the \chandra image was checked by comparing with the radio emission. The peak of counts at 3--7~keV, where a nuclear-emission was dominant \citep[Section~\ref{sec:ironmap}; e.g., ][]{Riv14,Kaw16b}, was found at (R.A., Dec. = 5h52m11.367s, $-$7d27m22.496s). This  differed only by $\approx$ 0.15 arcsec from the radio emission peak at (R.A, Dec.  = 05h52m11.377s, $-$7d27m22.492s) (see Section~\ref{sec:alm_ana}). Thus, no correction for the image was made.

\subsection{ALMA}\label{sec:almadata} 

We utilized the Band 6 ($\sim$ 270 GHz) ALMA data with the program ID = \#2012.1.00474.S (PI: N. Nagar). The data was taken with an on-source exposure of 2.2~ksec on 2015 March 14, when 37 antennas were operated. It covered CO($J$ = 2--1) emission at the rest-frame frequency of $\nu_{\rm rest}$ = 230.538~GHz. The maximum angular scale, which can be recovered from observations, was $\sim$ 8.6 arcsec ($\approx$ 1.38 kpc) from a minimum projected baseline of 19~m. This was larger than the central $\approx$ 1.3 kpc, or 8 arcsec, scale structure of interest. A spectral window that covered CO($J$ = 2--1) emission had a total bandwidth of 1.875 GHz with a central frequency of 228.792 GHz and 1920 channels. The raw spectral resolution was 0.98 MHz ($\approx$ 1.3 km s$^{-1}$), but eight spectral elements were binned to achieve a resolution of $\approx$ 7.8~MHz (10 km s$^{-1}$), as explained in Section~\ref{sec:alm_ana}. Standard flux, bandpass, and phase calibrations were made by observing Ganymede, J0423-0120, and J0541-0541, respectively. Detailed analyses of this data is described in Section~\ref{sec:alm_ana2}. 

\section{Data analysis}\label{sec:datana}

\subsection{\chandra data analysis}\label{sec:cxodatana}

In the former part of this subsection (Sections~\ref{sec:high-res} and \ref{sec:ironmap}), we describe imaging analyses and show extended X-ray emission. Then, we present spatially-resolved spectral analyses, confirming the extension. 

\begin{figure*}[!t]
  \centering
  \includegraphics[scale=0.23]{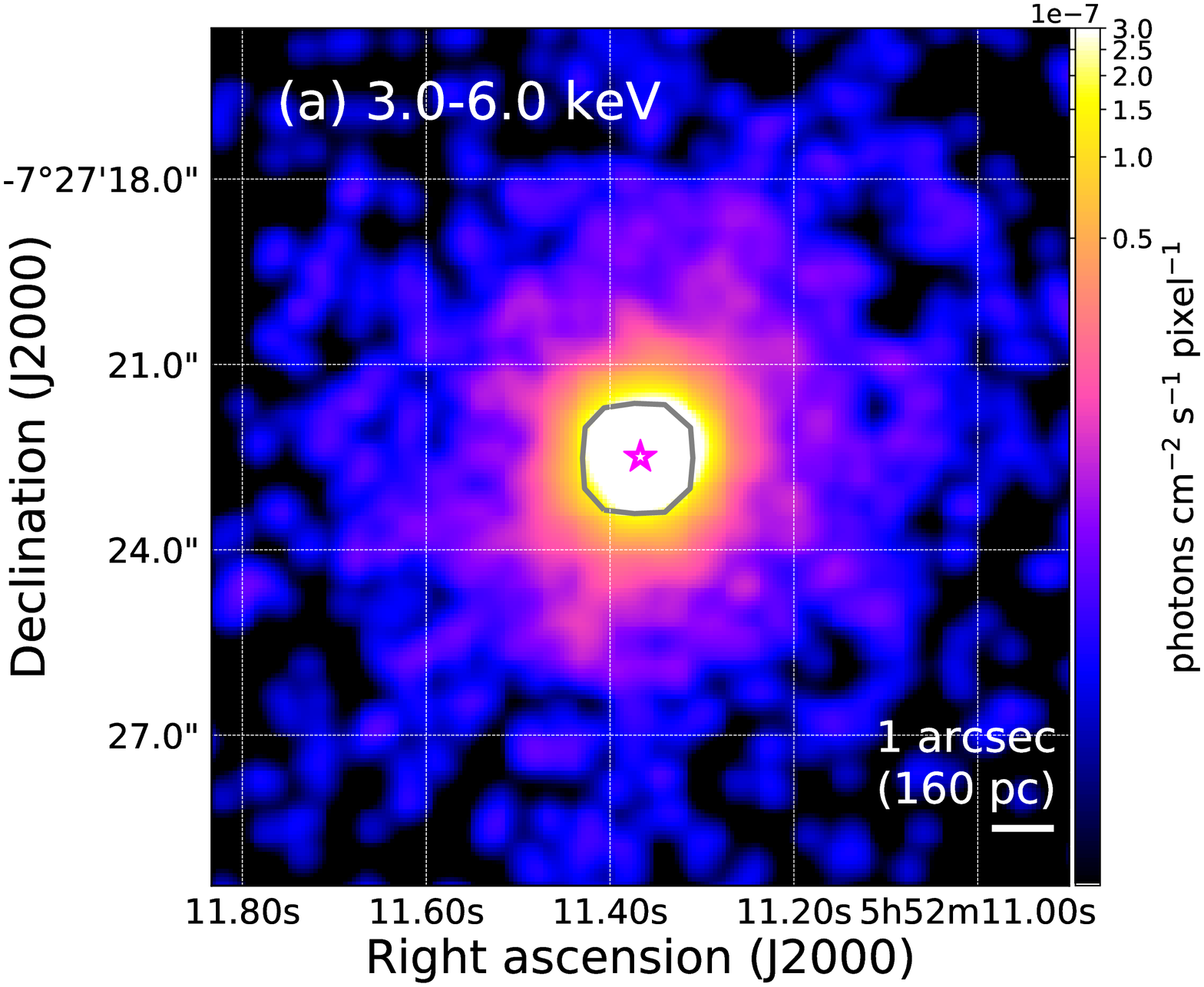}
  \includegraphics[scale=0.23]{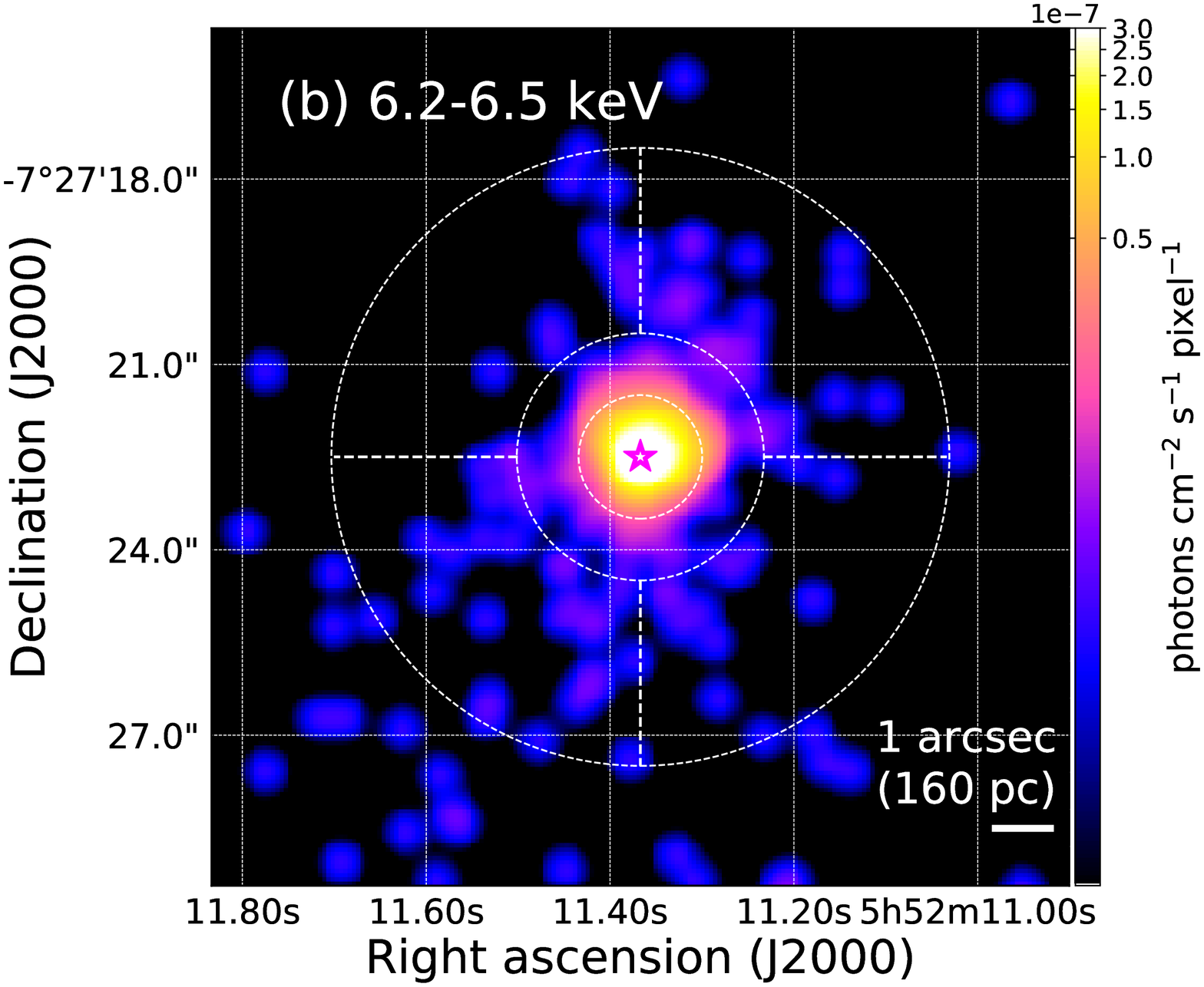}
  \includegraphics[scale=0.23]{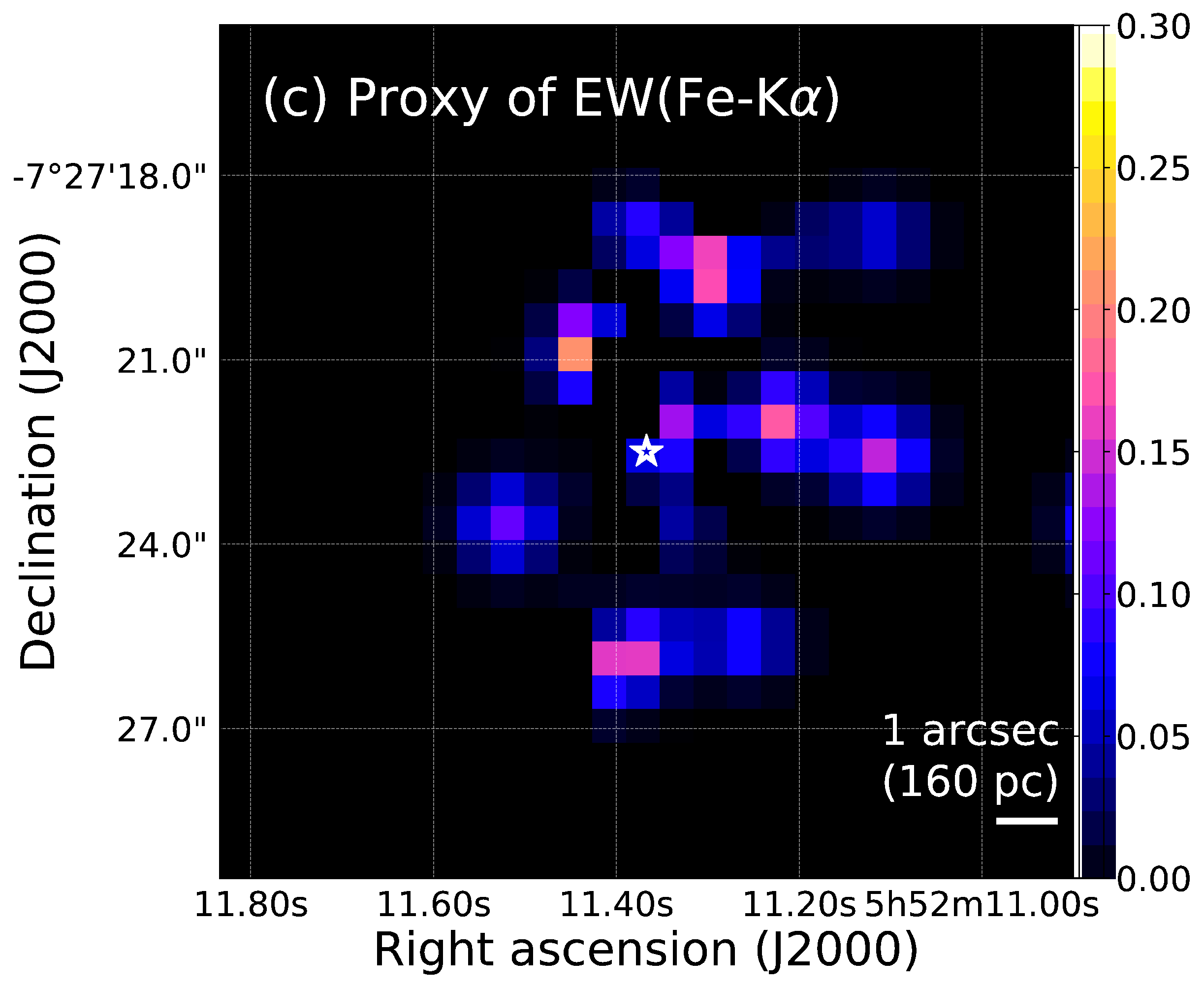} 
  \caption{\small{
    (a,b) X-ray images at 3.0--6.0 keV and 6.2--6.5 keV within the central $\approx$ 14$\times$14 arcsec$^2$ region. The size of each pixel is 0.0625$\times$0.0625 arcsec$^2$, and adaptive smoothing was performed with a Gaussian kernel with FWHM = 0.492 arcsec. In (a), the pixels with pile-up fractions larger than 5\% are enclosed by the gray solid line. In (b), the inner annulus between 1 and 2 arcsec was used to estimate the nuclear X-ray emission (Section~\ref{sec:ironmap}), and on the outside, the four fan-shaped regions between 2 and 5 arcsec with an angle of 90 degrees were defined to extract spectra of extended X-ray emission (Section~\ref{sec:xspe}). 
    (c) Ratios between the 6.2--6.5 keV and 3.0--6.0 keV images from which un-resolved nuclear emission was subtracted. The size of the region is the same as those of (a) and (b). Original images before calculating the ratios had a pixel size of 0.49$\times$0.49 arcsec$^2$. Adaptive smoothing with a Gaussian kernel with FWHM = 1.0 arcsec was made after the calculation. Each of the figures has a magenta or white star at the 3--7 keV peak.
    }
  }\label{fig:x_ima} 
\end{figure*}

\subsubsection{Observed high-angular resolution X-ray images}\label{sec:high-res} 

We show \chandra 3.0--6.0~keV and 6.2–-6.5~keV images in Figure~\ref{fig:x_ima},  respectively giving brief impressions about spatial distributions of continuum emission and the 6.4 keV Fe-K$\alpha$ line. They were made by sampling counts on a sub-pixel scale of 0.0615$\times$0.0615 arcsec$^2$, that is, by adopting the energy-dependent sub-pixel event-re-positioning algorithm  \citep[e.g.,][]{Tsu01,Mor01,Li03,Li04}. Then, we made smoothing via a Gaussian kernel with FWHM = 0.492 arcsec, corresponding to the ACIS CCD pixel size. Exposure maps for the softer and harder X-ray images were calculated at intermediate energies of 4.5 keV and 6.35 keV, respectively. At first glance, we can see 6.2--6.5 keV emission elongated in a northwest-southeast direction (Figure~\ref{fig:x_ima}(b)). Such can be seen also in the 3.0--6.0 keV image (Figure~\ref{fig:x_ima}(a)). 


\subsubsection{Spatial distribution of Fe-K$\alpha$ emission}\label{sec:ironmap} 

We revealed extended Fe-K$\alpha$ emission by removing unresolved nuclear emission, which spreads by a point spread function. The nuclear image was created by simulating observations where the nuclear emission only existed. Our simulations were made using a Monte-Carlo simulator of the MARX software (ver. 5.3.3; \citealt{Dav12}), which generates photons and projects them onto a detector plane while taking account of the mirror and detector responses. 

\begin{figure}[!h]
  \includegraphics[scale=0.5]{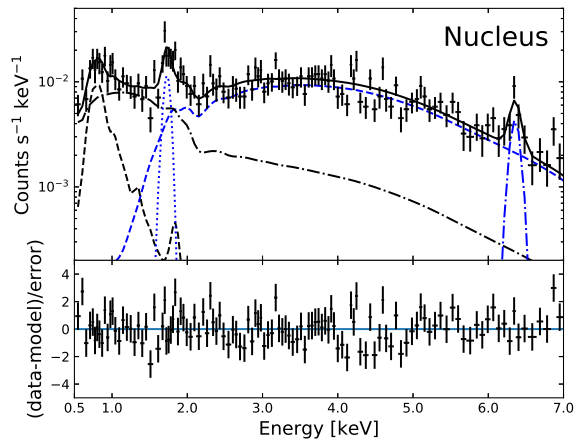}
  \caption{\small{
      X-ray spectrum (black crosses) extracted from an annulus between 1 and 2 arcsec and the best-fit model (black solid line). An absorbed power-law component (blue dashed line),  1.74 keV Si-K$\alpha$ line (blue dotted line), and 6.4 keV Fe-K$\alpha$ one (blue dot-dashed line) were assumed to come from the unresolved central X-ray source. Additionally, an un-absorbed power-law (black dot-dashed line) and the optically-thin thermal emission (\texttt{apec}; black dashed line) were fitted to mainly reproduce soft X-ray emission. 
      The lower data represent the residuals. 
    }
  }\label{fig:nuc_spe} 
\end{figure}

We determined the input nuclear emission spectrum by extracting its component from a spectrum in an annulus between 1 and 2 arcsec (Figure~\ref{fig:x_ima}(b)). Therein, the pile-up effect was negligible.  A background spectrum was estimated from a blank 50 arcsec radius circle located in the same CCD. The 0.5--7.0 keV spectrum was fitted with absorbed and un-absorbed power-law components, two Gaussian functions (\texttt{zgauss} in XSPEC terminology) for the 1.74 keV Si-K$\alpha$ and 6.4 keV Fe-K$\alpha$ lines, and an optically-thin thermal component  (i.e., the \texttt{apec} model in XSPEC). By considering poor photon statistics, the photon indices of the power-law components were fixed to 1.65, a canonical value suggested from past studies of NGC~2110  \citep[e.g., ][]{Eva07,Kaw16b}. Also, because of the CCD poor energy resolution, we fixed the line widths at 0.1 eV. Adopting a width smaller than those constrained by \chandra grating observations \citep[i.e., $\approx$ 2300 km s$^{-1}$; ][]{Shu10} did not affect our result. The best-fit model was determined based on the $C$-statistic \citep{Cas79}, appropriate even for low photon statistics. Goodness of fit was examined by following the procedure given in \cite{Kaa17}. The expected $C$-statistic value ($C_{\rm exp}$) and variance ($C_{\rm var}$) from a model was compared with an observed value ($C_{\rm obs}$). A model would be acceptable at the 90\% confidence level if $C_{\rm obs} < C_{\rm exp} + 1.28\times \sqrt{C_{\rm var}}$ \citep{Kaa17}. Eventually, we obtained the best-fit model with $C_{\rm obs}$/$C_{\rm exp}$/$C_{\rm var}$ = 356/472/29. 
\modtext{A good agreement between the data and best-fit model 
is found in Figure~\ref{fig:nuc_spe}. The resultant parameters are summarized in Table~\ref{tab:xnuc}. An observed flux at 2--10 keV, where the absorbed power-law component is dominant, is $2.3\times10^{-11}$ \ergcms, and the intrinsic luminosity of the absorbed power-law component at 2--10 keV was $4\times10^{42}$ \ergs, which is within those measured previously \citep[e.g., ][]{Mar15}. The equivalent width of the Fe-K$\alpha$ line was 152$^{+141}_{-107}$ eV, consistent with those previously measured for NGC 2110 by \cite{Mar15} as well as those typical in moderately obscured AGNs \citep[e.g., 30--500~eV; ][]{Kaw16b}.}



\begin{deluxetable}{cccc}
\tabletypesize{\footnotesize}
\tablecaption{Best-fit parameters of the annulus spectrum \label{tab:xnuc}}
\tablewidth{0pt} 
\startdata \vspace{-0.1cm} \\ 
   & Parameter & Best-fit  & Units \\ \hline 
  (1) & $N_{\rm H}$ & $4.0^{+0.7}_{-0.6}$  & 10$^{22}$ cm$^{-2}$ \\
  (2) & $\Gamma$    & 1.65 & \\
  (3) & $N^{\rm abs}_{\rm PL}$ &  $6.3^{+0.6}_{-0.5}$ 
  & 10$^{-3}$ photons keV$^{-1}$ cm$^{-2}$ s$^{-1}$\\
  (4) & $N^{\rm un-abs}_{\rm PL}$ &  $0.51\pm0.14$  
  & 10$^{-3}$ photons keV$^{-1}$ cm$^{-2}$ s$^{-1}$\\
  (5) & $kT$ &  $0.57^{+0.18}_{-0.23}$ & keV \\
  (6) & $N_{\rm apec}$ &  $1.8^{+0.09}_{-0.08}$ & $10^{-4}$ \\
  (7) & $N_{\rm Fe-K{\alpha}}$ &  $46^{+26}_{-22}$ & 10$^{-6}$ photons cm$^{-2}$ s$^{-1}$ \\
  (8) & $N_{\rm Si-K{\alpha}}$ &  $69^{+19}_{-18}$ & 10$^{-6}$ photons cm$^{2}$ s$^{-1}$ \\    
  (9)  & $F_{\rm 2-10~keV}$ & 2.3 & 10$^{-11}$ erg cm$^{-2}$ s$^{-1}$\\ 
  (10) & $L_{\rm 2-10~keV,PL}$ & 3.7 & 10$^{42}$ erg s$^{-1}$\\ 
  (11) & $C_{\rm obs}$/$C_{\rm exp}$/$C_{\rm var}$ & 356/472/29 \\ 
\enddata
\tablecomments{
Rows:
(1) Hydrogen column density in the sightline.  
(2) Photon index of the absorbed and un-absorbed power-law components. 
The value was fixed. 
(3,4) Normalizations of the two power-law components at 1 keV.  
(5,6) Temperature and normalization of the \texttt{apec} model (see https://heasarc.gsfc.nasa.gov/xanadu/xspec/manual/node134.html for more details). 
(7,8) Normalizations of the Gaussian models (\texttt{zgauss}) for the Fe-K$\alpha$ and Si-K$\alpha$ emission. 
\modtext{(9) Observed flux at 2--10 keV, where the absorbed power-law component is dominant.
(10) Intrinsic luminosity of the absorbed power-law component at 2--10 keV.} 
(11) $C$-statistic values. 
}
\end{deluxetable}

From the best-fit spectral components, we defined the sum of the absorbed power-law and two emission lines as the nuclear emission. The other components (i.e., the un-absorbed power-law emission and the thermal emission) would largely include in-situ emission given that they were un-absorbed. Indeed, if a component is from the nucleus, it should be absorbed by the sightline obscuration. 

\begin{figure*}
  \centering
  \includegraphics[angle=-90,scale=0.3]{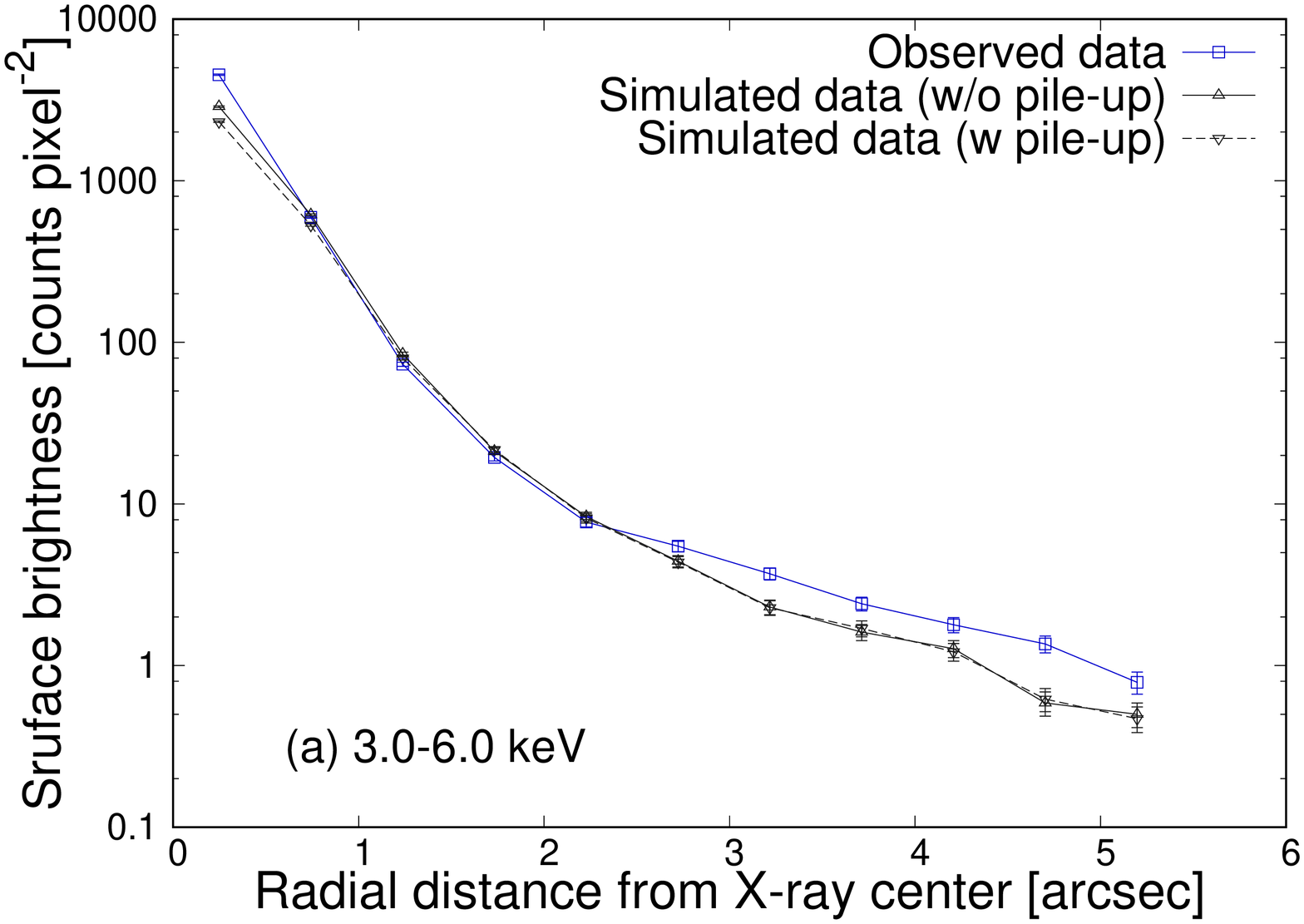}
  \includegraphics[angle=-90,scale=0.3]{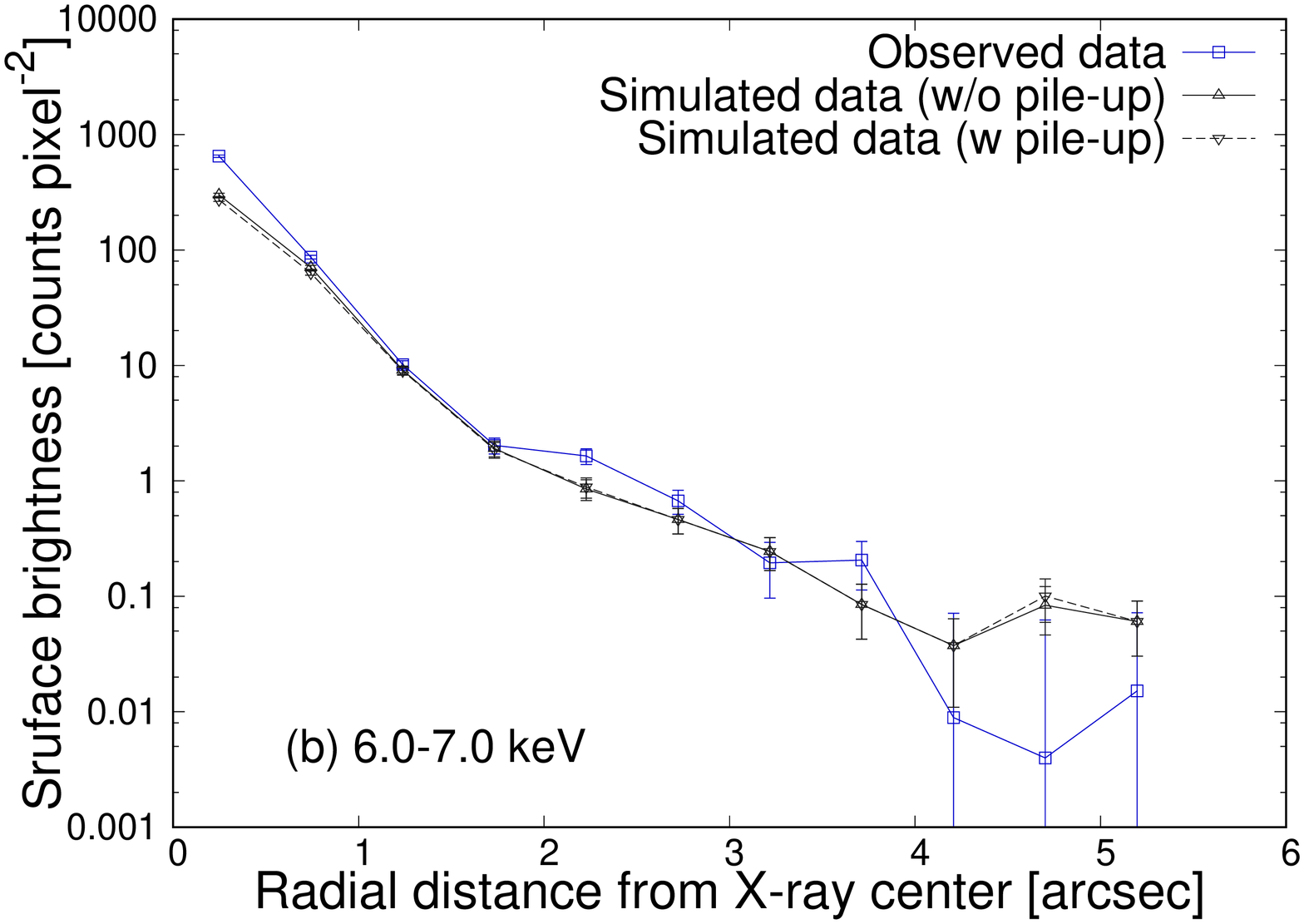} \vspace{.5cm}
  \caption{\small{
  (a) Comparison between observed (blue) and simulated (black) radial profiles of the surface brightness in units of counts pixel$^{-2}$ at 3.0--6.0 keV. The black solid and dashed lines show those without and with the pile-up effect, respectively.  
  (b) Same as (a), but for the 6.0--7.0 keV band. 
    }
  }\label{fig:xr_comp} 
\end{figure*}

With the defined spectrum, we simulated \chandra imaging observations. The source position was set to the 3--7 keV peak (magenta stars in Figure~\ref{fig:x_ima}). The other parameters necessary for the simulation were set so as to reproduce the ObsID = 883 observation. To mitigate uncertainty of each Monte-Carlo simulation result, we created 1000 images, and took their average. 

The validity of our simulation was examined by comparing the radial profiles of the observed and simulated images. Figure~\ref{fig:xr_comp} shows broad consistency between the profiles, or the images. Although a factor of $\sim$ 2 difference was seen in the innermost pixels, this would be due to incompleteness of the simulation software as was  remarked by the developers\footnote{https://space.mit.edu/cxc/marx/tests/index.html}\footnote{As a possibility for the discrepancy, one might suggest that we missed a fraction of the nuclear emission. For example, the inclusion of the other components (i.e., the un-absorbed power-law emission and/or the thermal emission) was however not preferred. If we take them into consideration, the simulation yields a profile larger than the observed one at larger radii. Another idea was that our simulation underestimated photons which were produced as the result of piled softer photons. However, if we consider a more soft-X-ray luminous spectrum, the simulated profile largely exceeds the observed one below 3~keV.}. Because the simulated profile did not exceed the observed one significantly at any radii and energies, we concluded to proceed with the above simulation results.


At last, in Figure~\ref{fig:x_ima}(c) we show the ratios between 6.2--6.5 keV and 3.0--6.0 keV images from which the nuclear emission was subtracted. The ratio is a proxy of the equivalent width (EW) of the Fe-K$\alpha$ line. The X-ray photons were sampled at a larger pixel size of 0.49$\times$0.49 arcsec$^2$ to increase S/N ratios. Particularly in the 3.0--6.0 keV image, we excluded pixels with counts smaller than 2, corresponding to $\approx$ 1$\sigma$ level for the mean value of 0 counts \citep{Geh86}. This was made to avoid extreme ratios due to quite low photons in the 3.0--6.0 keV image. Exposure maps were created in the same way as for the previous images. The image of the ratios was smoothed with the Gaussian kernel with FWHM = 1.0 arcsec. We can see a structure extending preferentially in a northwest-southeast direction out to $\approx$ 3~arcsec, or $\sim$ 500~pc, on each side. We caution that the ratio image changes depending on the significance criterion imposed on the 3.0--6.0 keV image. However, the extended morphology can be seen generally in most cases. Also, we emphasize that the result is consistent with the model-independent image of Figure~\ref{fig:x_ima}(b) and is confirmed by the spectral analysis below.

\subsubsection{Spatially resolved X-ray spectral analysis}\label{sec:xspe}

We confirmed the un-isotropic extended Fe-K$\alpha$ emission suggested from  Figure~\ref{fig:x_ima} by spatially-resolved spectral analysis. We focused on four fan-like sub-regions, outlined in Figure~\ref{fig:x_ima}(b). They were referred to as NW, SW, SE, and NE, depending on their directions from the nucleus. Starting from north, the fan-like regions were defined to have an angle of 90 degrees. Their inner and outer radii were 2 arcsec and 5 arcsec, respectively. A background spectrum was estimated from a blank 50 arcsec radius circle located in the same CCD. The spectra were binned so as to have at least one count at each bin. Two different types of response files for the in-situ emission and the nuclear emission were generated using the CIAO tool \texttt{specextract}.

\begin{figure*}[!h]
  \centering
  \includegraphics[scale=0.46]{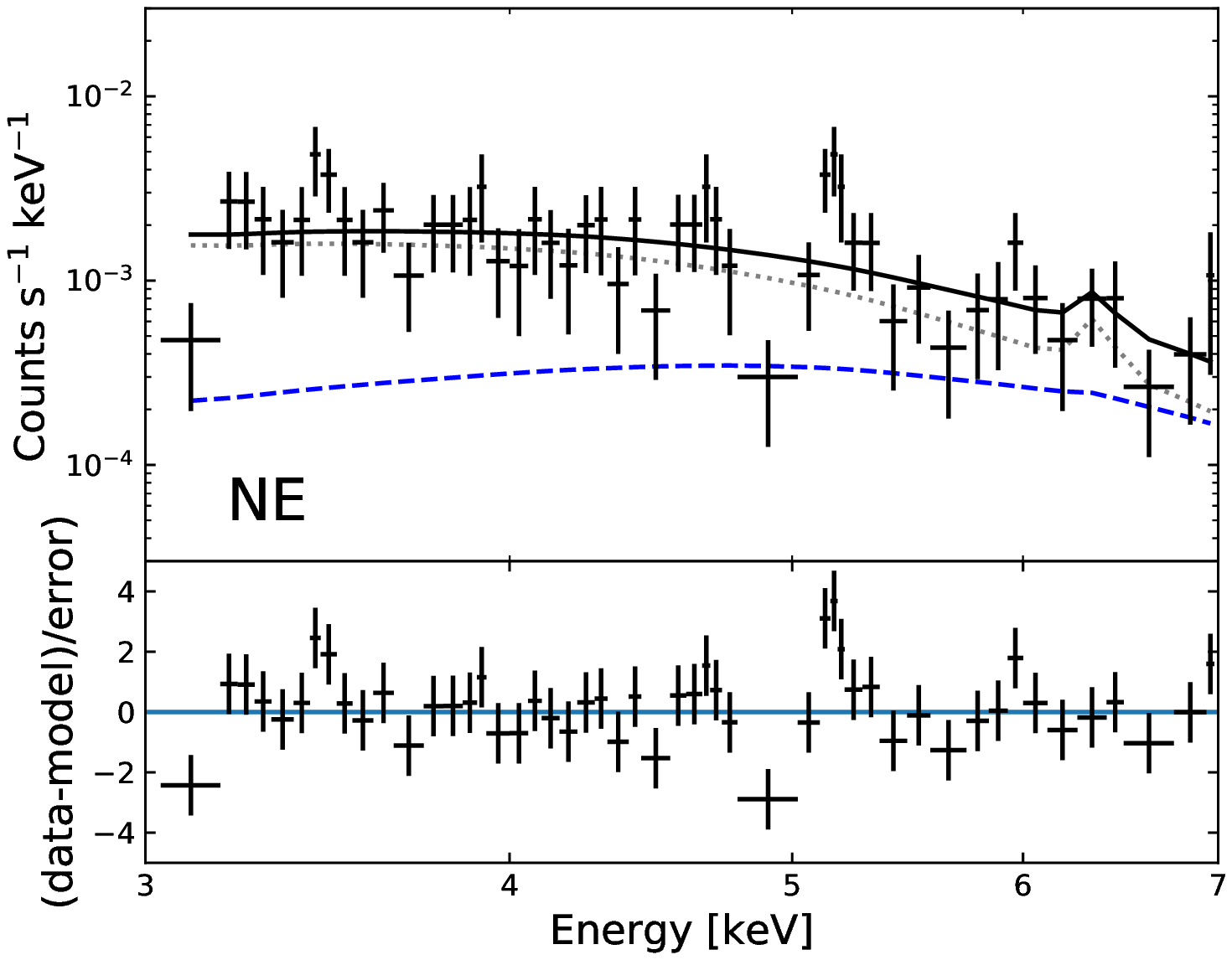}
  \includegraphics[scale=0.46]{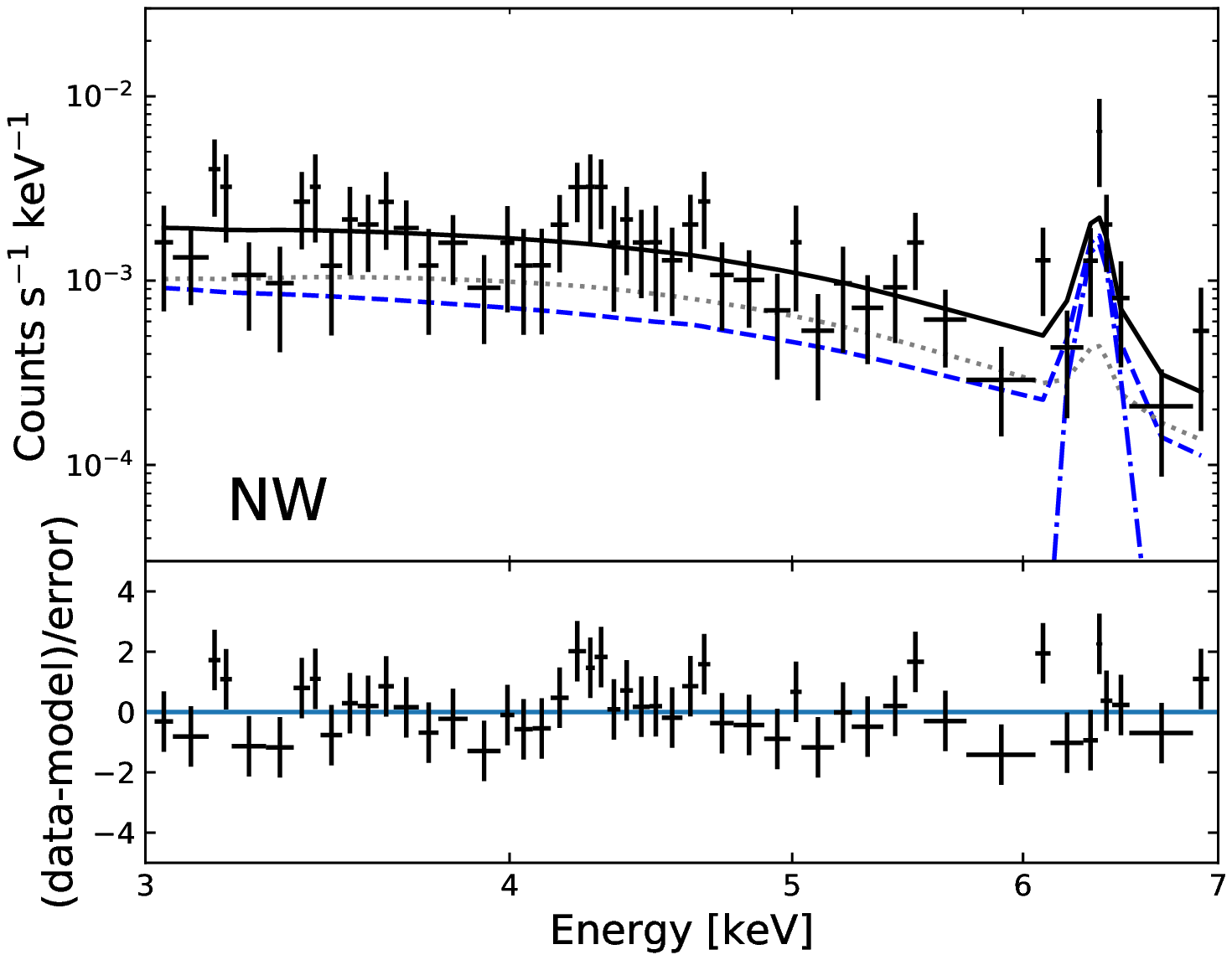} \\
  \includegraphics[scale=0.46]{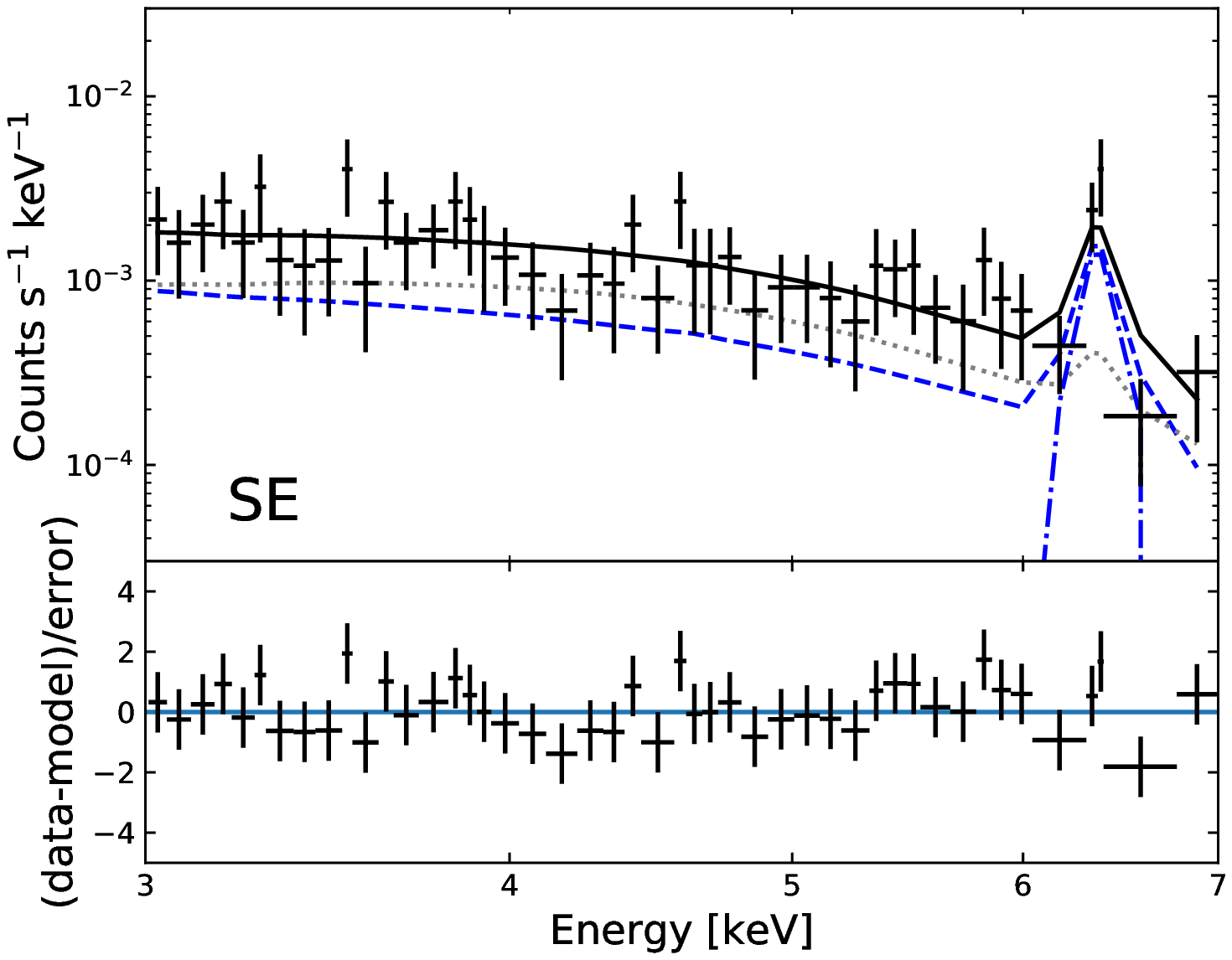}
  \includegraphics[scale=0.46]{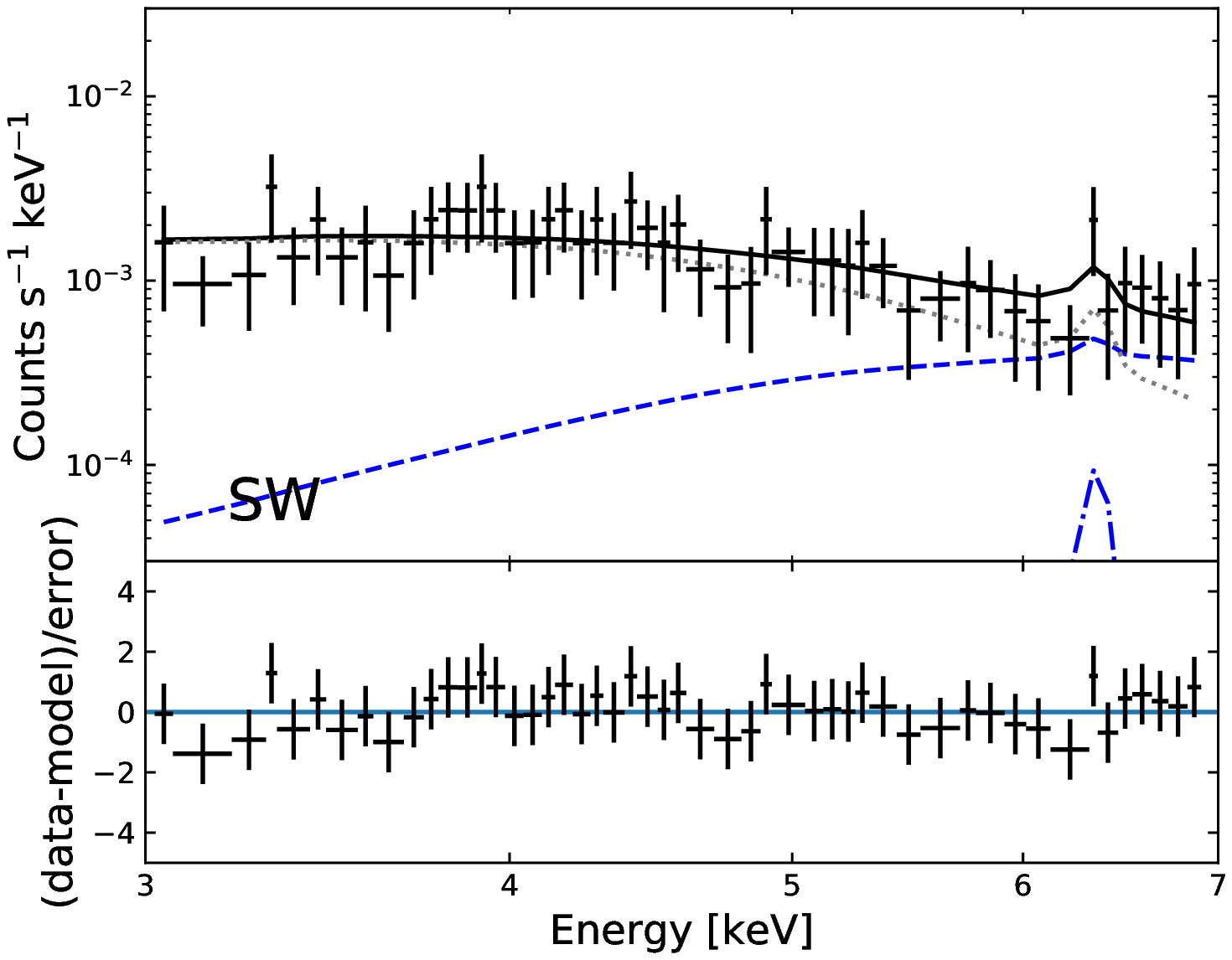} 
  \caption{\small{
      X-ray spectra (black crosses) extracted from the four fan-like regions defined in Figure~\ref{fig:x_ima}(b). The best-fit models (solid line) were determined by fitting the contaminating X-ray emission from the nuclear point source (dotted gray line) and the  extended emission of interest (blue dashed line), consisting of a power-law and Fe-K$\alpha$ emission (blue dot-dashed line). The lower data represent the residuals. 
    }
  }\label{fig:x_spe} 
\end{figure*}

While taking account of the nuclear emission (the sum of the blue lines in Figure~\ref{fig:nuc_spe}) as a fixed component, we fitted the spectra with a power-law and a Gaussian function, reproducing continuum emission and the Fe-K$\alpha$ line, respectively. The line width was fixed at 0.1 eV. Therefore, the normalization and photon index of the power-law component and the normalization of the Gaussian function were left as free parameters. The spectra folded by the response functions and their best-fit models are shown in Figure~\ref{fig:x_spe}. The parameters are summarized in Table~\ref{tab:best_pow}.

\begin{deluxetable*}{ccccccccccccccc}[!t]
\tabletypesize{\scriptsize}
\tablecaption{Best-fit parameters of the four sub-regions spectra with the power-law and Gaussian models \label{tab:best_pow}}
\tablewidth{0pt} 
\startdata \vspace{-0.1cm} \\ 
      &              & NE & NW & SE & SW \\ \hline 
  (1) & $\Gamma$     & $-1.3^{+1.4}_{-1.6}$
                     & $1.0^{+0.9}_{-0.8}$
                     & $1.2^{+1.0}_{-0.9}$
                     & $-4.1^{+1.9}_{-3.3}$ \\
  (2) & $N_{\rm PL}$ & $0.12^{+0.96}$
                     & $7.0^{+15.4}_{-4.9}$ 
                     & $8.0^{+21.4}_{-5.9}$ 
                     & $0.0010^{+0.0253}$ 
                     & 10$^{-6}$ photons keV$^{-1}$ cm$^{-2}$ s$^{-1}$\\
  (3) & $N_{\rm Fe-K{\alpha}}$ 
                     & $0.12^{+4.86}$ 
                     & $16^{+7}_{-6}$ 
                     & $15^{+7}_{-6}$ 
                     & $0.96^{+5.28}$
                     & 10$^{-7}$ photons cm$^{-2}$ s$^{-1}$ \\
  (4) & EW$_{\rm Fe-K{\alpha}}$ 
                     & ---   
                     & 1.5$^{+2.1}_{-1.0}$ 
                     & 1.6$^{+2.6}_{-1.1}$ 
                     & ---  
                     & keV \\ 
  (5) & $C_{\rm obs}$/$C_{\rm exp}$/$C_{\rm var}$
                     & 126/160/14
                     & 105/154/14 
                     & 104/143/13 
                     & 101/161/15 \\                     
\enddata
\tablecomments{
Rows:
(1,2) Photon index and normalization at 1 keV of the power-law component. 
(3) Normalization of the Gaussian model (\texttt{zgauss}), reproducing the Fe-K$\alpha$ emission. 
(4) Equivalent width of the Fe-K$\alpha$ emission. 
(5) $C$-statistic values. 
}
\end{deluxetable*}

An important point is the significant detection of the Fe-K$\alpha$ emission lines only in the NW and SE regions.
The difference in $C_{\rm obs}$ between the models with and without the line was found $\sim$ 12 in the NW and SE cases, suggesting the high significance. This was not the case in the NE and SW regions. The ratio between the sum of the NW and SE extended components and the unresolved nuclear one was estimated to be $\sim$ 7\%. \modtext{The EWs in the NW and SE regions were found 1.5--1.6 keV. One might be confused by the EWs, because such high EWs were observed often in Compton-thick AGNs \citep{Ric15,Tan18} but NGC 2110 hosts just a lightly obscured AGN with $N_{\rm H} \sim 10^{22}$ cm$^{-2}$ \citep{Mar15}. As later detailed in Section~\ref{sec:ext_xray}, the high EWs can be because we just see only reflected emission  without transmitted emission. In this case, such high EWs can be reproduced with various column densities \citep[e.g., see the right panel of Figure 14 of][]{Ike09}. This is natural given that we focus solely on the extended components. Instead, the observed EW from the much brighter nuclear component ($\approx 150$ eV in Section~\ref{sec:ironmap} and Figure~\ref{fig:nuc_spe}) is better to be compared with those generally obtained, or those from spatially \textit{un-resolved} spectra. In the case, that is consistent with those of moderately obscured AGNs with $N_{\rm H} = 10^{22-24}$ cm$^{-2}$  \citep[e.g., 30--500~eV; ][]{Kaw16b}.}



We further fitted a reflection model of the Ikeda torus model \citep{Ike09} to the spectra in the NW and SE regions, where the Fe-K$\alpha$ emission was significantly detected. By this analysis, it was possible to roughly estimate a column density along the propagating path of the nuclear X-ray emission. The torus model calculated spectra from a spherical structure with a bi-conical gas-free region irradiated by a central point source with a power-law spectrum. The photon index of the power-law spectrum was fixed to 1.65  \citep[e.g., ][]{Eva07,Kaw16b}. The half opening angle and our inclination angle from the polar axis were by default set to 60 degrees and 30 degrees. Different choices of the parameters did not affect our conclusion.  Free parameters left were the normalization of the power-law component and the column density in the equatorial plane. By adopting the $C$-statistic method, we got acceptable fits (i.e., $C_{\rm obs} < C_{\rm exp} + 1.28 \times \sqrt{C_{\rm var}}$) for both the NW and SE spectra. The obtained parameters are summarized in Table~\ref{tab:best_torus}. The column densities were found $\sim 10^{23}$ cm$^{-2}$ at most. 

\begin{deluxetable*}{ccccccccccccccc}[!b]
\tabletypesize{\scriptsize}
\tablecaption{Best-fit parameters of the two sub-regions spectra with the torus models \label{tab:best_torus}}
\tablewidth{0pt} 
\startdata \vspace{-0.1cm} \\ 
      &              & NW & SE \\ \hline 
  (1) & $\Gamma$     & 1.65 & 1.65 & \\ 
  (2) & $N_{\rm PL}$ & $1.5^{+5.0}_{-0.8}$
                     & $1.9^{+4.3}_{-1.2}$ 
                     & 10$^{-3}$ photons keV$^{-1}$ cm$^{-2}$ s$^{-1}$\\
  (3) & $N^{\rm eq}_{\rm H}$  
                     & $4.6^{+6.8}$ 
                     & $3.0^{+7.2}$ 
                     & 10$^{22}$~cm$^{-2}$ \\ 
  (4) & EW$_{\rm Fe-K{\alpha}}$ 
                     & 1.6 
                     & 1.9
                     & keV \\ 
  (5) & $C_{\rm obs}$/$C_{\rm exp}$/$C_{\rm var}$
                     & 105/154/14
                     & 105/143/13 \\
\enddata
\tablecomments{
Rows:
(1,2) Photon index and normalization at 1 keV of the power-law component. 
The photon index was fixed. 
(3) Hydrogen column density in the equatorial plane. 
(4) Equivalent width of the Fe-K$\alpha$ line. 
(5) $C$-statistic values. 
}
\end{deluxetable*}

\subsection{ALMA data analysis}\label{sec:alm_ana2}

In the following two subsections, first we show basic pieces of information on  the CO($J$ = 2--1) line, and then we present its geometrical and kinematic properties. 

\subsubsection{Basic properties of CO($J$ = 2--1) emitting gas}\label{sec:alm_ana} 

\begin{figure*}[!t]
  \centering
  \includegraphics[scale=0.23]{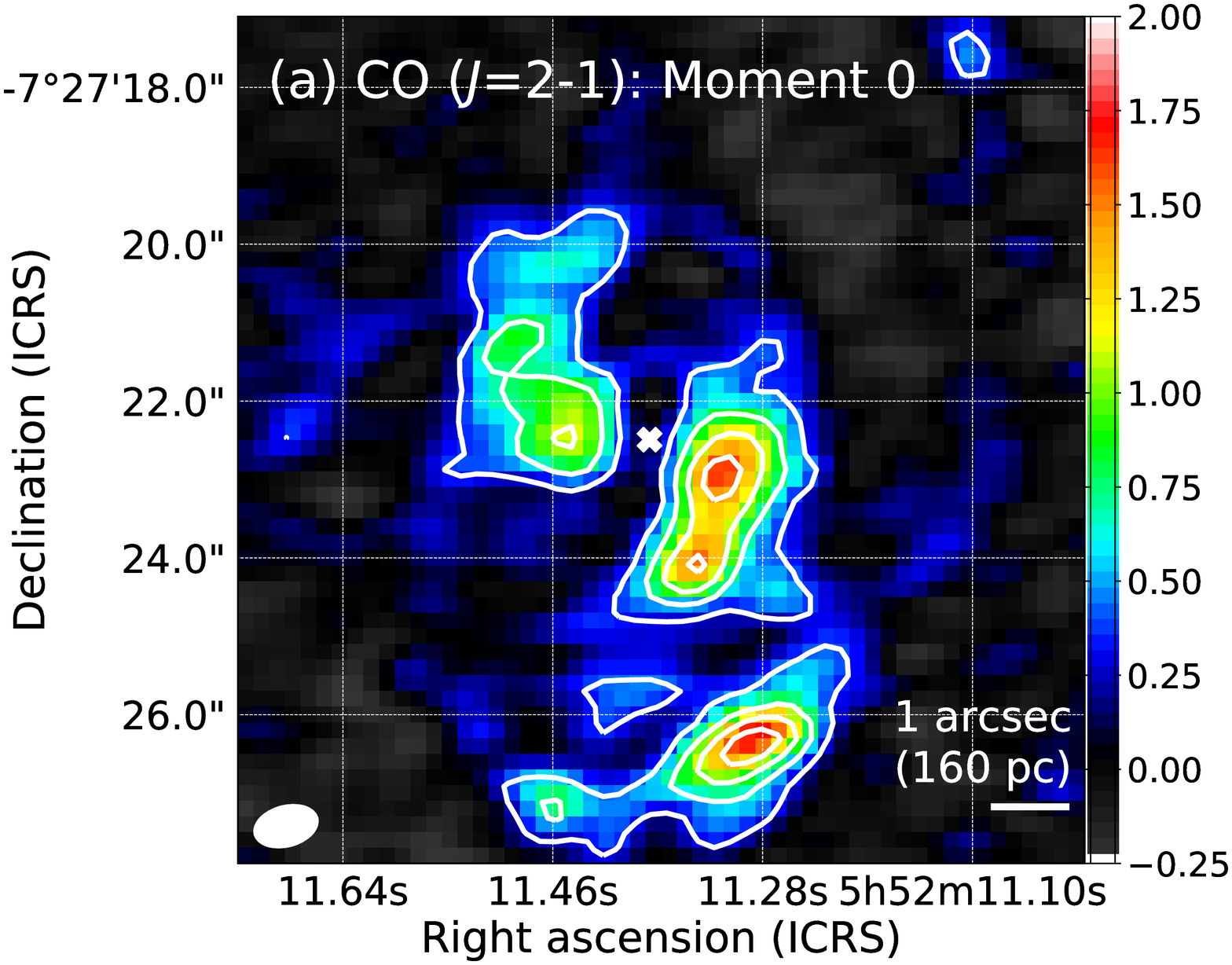}
  \includegraphics[scale=0.23]{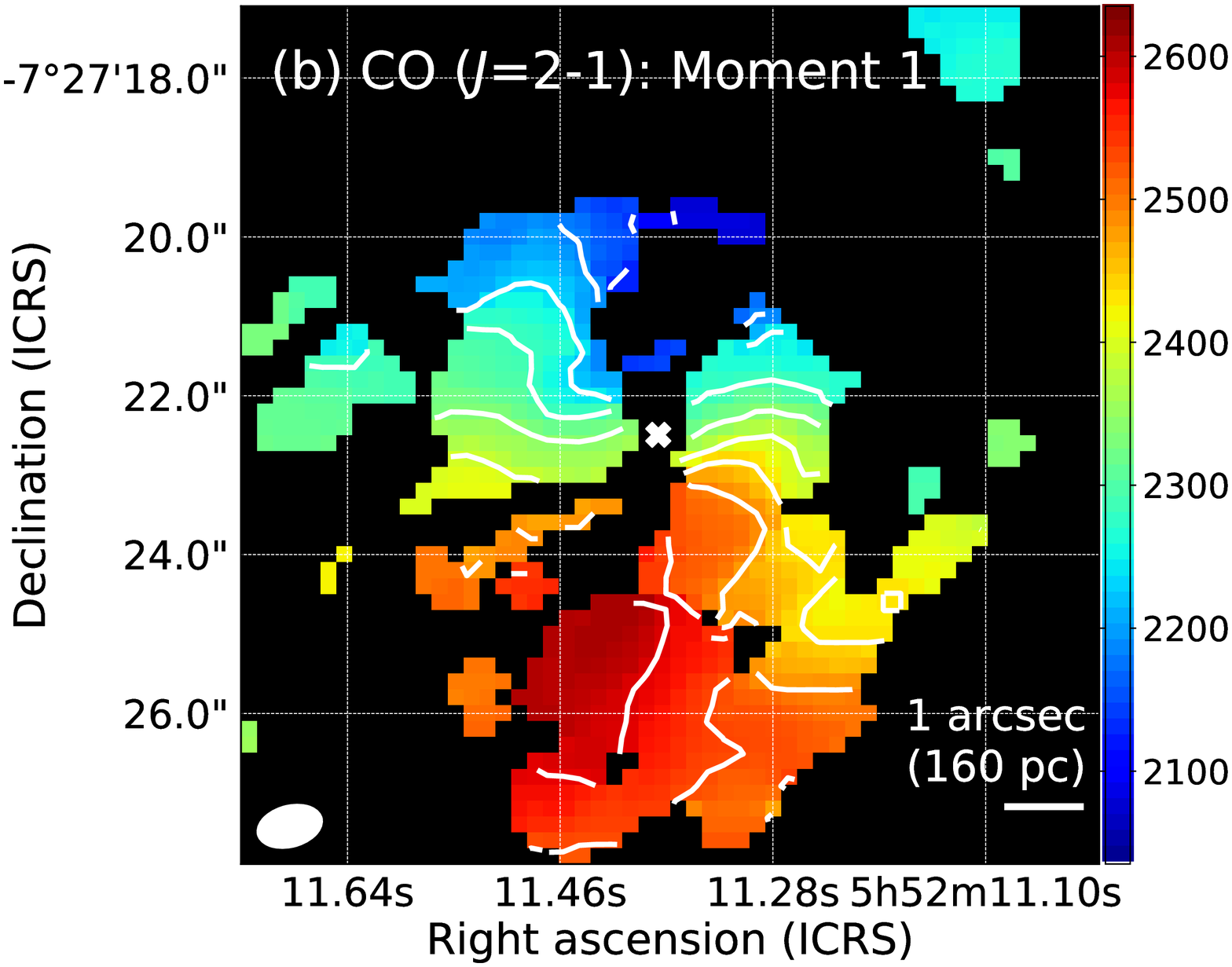}
  \includegraphics[scale=0.23]{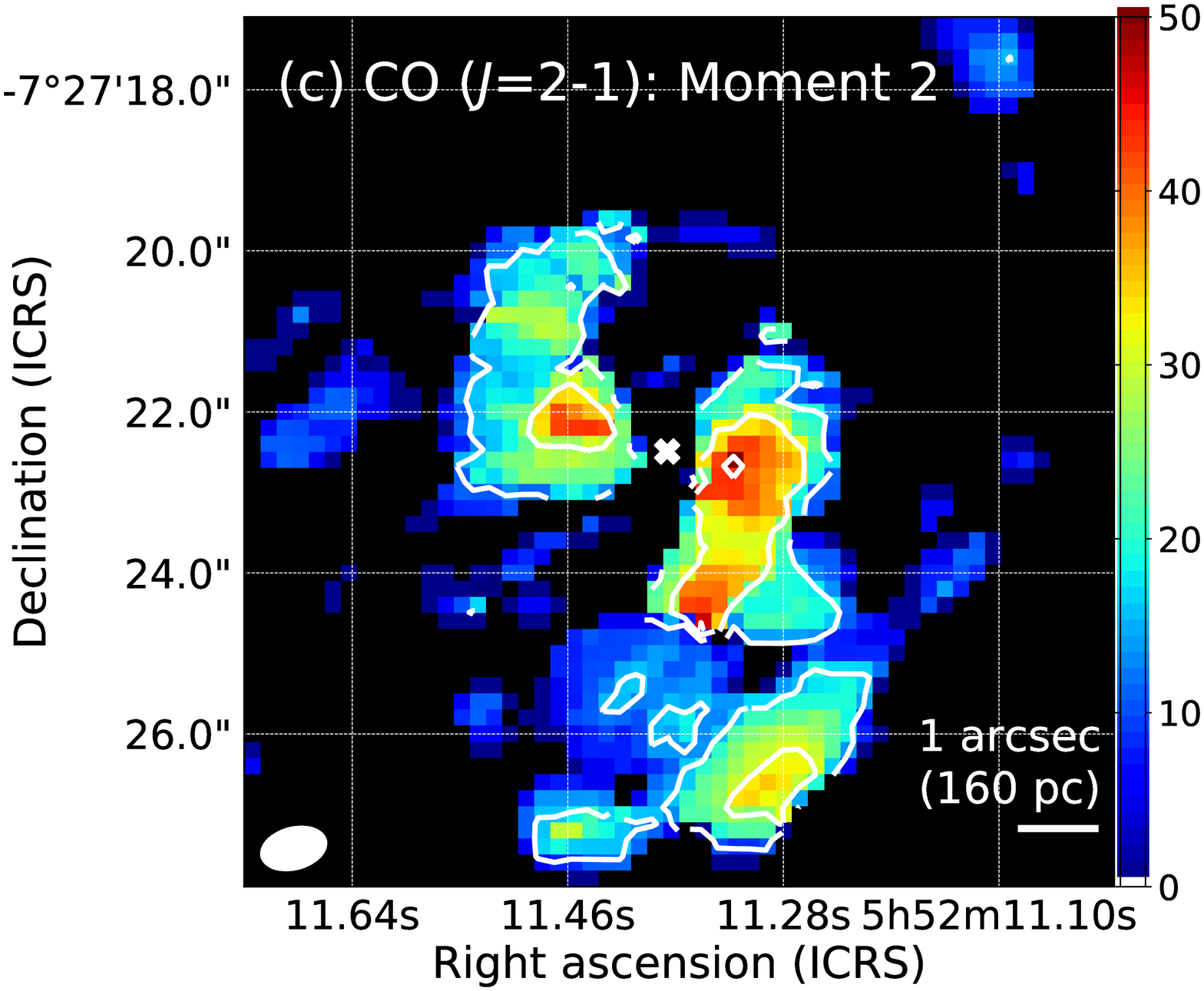}
  \caption{\small{
      (a) Moment 0 map of the CO($J$ = 2--1) line in units of Jy beam$^{-1}$ km s$^{-1}$. This was created in the range $V_{\rm LSR}$  = 1900--2800 km~s$^{-1}$. The contour levels are 5$\sigma$, 10$\sigma$, 15$\sigma$, and 20$\sigma$, where $\sigma$ is 0.073 Jy beam$^{-1}$ km s$^{-1}$. 
      (b) Moment 1 map in units of km s$^{-1}$, where the contours represent the $V_{\rm LSR}$  with steps of 50 km s$^{-1}$ from 2086 km~s$^{-1}$.
      (c) Moment 2 map in units of km s$^{-1}$ with the contours separated by 10 km s$^{-1}$. The beam size (bottom-left filled ellipse) is 0.52$\times$0.83 arcsec$^2$ with PA = $-$75.17 degrees. 
      In all figures, the region is the central  $\approx$11$\times$11 arcsec$^2$, and the white cross is located at the center of the continuum emission (R.A., Dec. =  5h52m11.377s, $-$7d27m22.492s).
    }
  }\label{fig:mol_map} 
\end{figure*}

\begin{figure}
  \centering
  \includegraphics[scale=0.3]{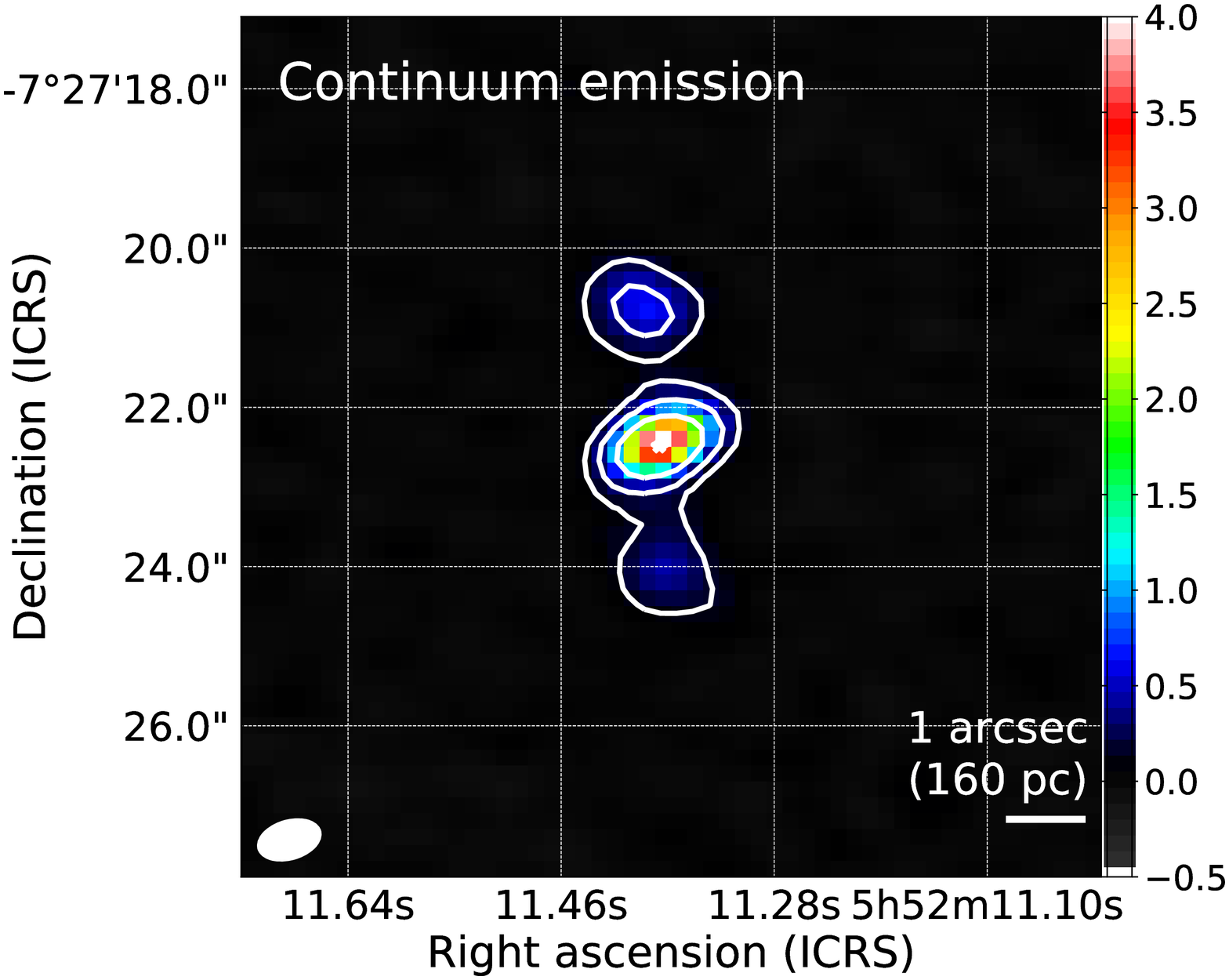}  
  \caption{\small{
      Continuum intensity map in units of mJy beam$^{-1}$ with contours of 5$\sigma$, 15$\sigma$, and 45$\sigma$, and 135$\sigma$, where $\sigma$ is 0.030 mJy beam$^{-1}$. The beam size (bottom-left filled ellipse) is 0.49$\times$0.81 arcsec$^2$ with PA = $-$73.80 degrees. The white cross is located at the center of the continuum emission (R.A., Dec. =  5h52m11.377s, $-$7d27m22.492s). 
    }
  }\label{fig:cont_map} 
\end{figure}

The ALMA data was reduced via the Common Astronomy Software Applications (CASA) \citep{McM07} with ver. 4.2.2, the same as in the Quality Verification by the ALMA Regional Center, and then was analyzed via the CASA with ver. 5.1.1. To extract the data of CO($J$ = 2--1) emission, continuum level was determined by fitting line-free channels with the 1st-ordered function, and was subtracted from the data cube via the task \texttt{uvcontsub}. The product was then deconvolved by the \texttt{clean} task with the Briggs-weighting with robust = 0.5 and gain = 0.1.
The achieved synthesized beam was $0.52\times0.83$ arcsec$^2$ with P.A. = $-75.17$ degrees. The channels were binned so that the velocity resolution was $\approx$ 10 km s$^{-1}$. The primary beam image correction was made via \texttt{impbcor}. The final continuum-subtracted data had a RMS noise of 0.76 mJy beam$^{-1}$. This was estimated from the CO emission-free channels. The moment 0, 1, and 2 maps of the CO($J$ = 2--1) line were made, and are shown in Figure~\ref{fig:mol_map}. The zeroth moment was calculated over the $V_{\rm LSR}$  range of 1900--2800 km s$^{-1}$. The other moment maps were calculated with 5$\sigma$ clipping in the same $V_{\rm LSR}$  range. Continuum emission was also mapped by the \texttt{clean} task (Figure~\ref{fig:cont_map}). The achieved synthesized beam was $0.49\times0.81$ arcsec$^2$ with P.A. = $-73.80$ degrees. The center of the nuclear component was estimated to be (R.A., Dec. = 5h52m11.377s, $-$7d27m22.492s) by fitting a single Gaussian function through the \texttt{imfit} task.



\subsubsection{Geometrical structure and kinetic properties of CO gas}\label{sec:kin}

We revealed a three-dimensional structure of the CO($J$ = 2--1) emitting gas disk to finally estimate a molecular gas volume density, essential to discuss the ionization state of the molecular gas. We reconstructed the structure by fitting a model of concentric tilted rings to the observed velocity field. Then, the rotation velocity and velocity dispersion were available from the model. With these, we derived the scale heights, or the thicknesses of the rings. Combined with an observed gas surface density, those eventually permitted an estimate of the molecular gas volume density, used to calculate  ionization parameters in Section~\ref{sec:xdr}. 


A fit of the concentric titled rings to the observed CO($J$ = 2--1) data was made by adopting the 3D Barolo code \citep{Dit15}. This code takes account of beam-smearing of an intrinsic structure. The main seven parameters are the dynamical center, systemic velocity ($V_{\rm sys}$), thickness of the ring, rotation velocity ($V_{\rm rot}$), velocity dispersion ($\sigma_{\rm dis}$), angle between the polar axis and our sightline ($i$), and P.A. ($\phi$), which can be determined in each of the rings. The P.A. is defined as an angle from north to the major axis of the receding half component in the anti-clockwise direction. The first three parameters were fixed. The dynamical center was set to the radio continuum one, as determined in Section~\ref{sec:alm_ana}. The systematic velocity was set to 2335~km~s$^{-1}$. The thickness of each ring was fixed at 0.5 arcsec ($\approx$ 80 pc) to be consistent with scale heights derived from the resultant rotation velocity and velocity dispersion. We then fitted  the remaining four parameters by starting from initial guesses. We set $i =$ 65~degrees as the initial guess in all the rings. A similar value was adopted by \cite{Ram19}, who also studied a kinematic structure of CO($J$ = 2--1) in NGC~2110. The adopted angle is larger than $i =$ 53 degrees, which was inferred from an observed aspect ratio between the major and minor axes of a  H$\alpha$+[N~II] emission distribution \citep[e.g.,][]{Wil85} and was sometimes adopted in past studies \citep[e.g., ][]{Gal99}. It is however not necessary to follow it, given that the H$\alpha$+[N~II] distribution may be disturbed by the nuclear emission. Also, if the thickness of the morphology cannot be ignored, a larger angle is more plausible. Finally, our choice ($i$ = 65 degrees) was made because we obtained a rotation curve that monotonously increased with radius, whereas we were not able to obtain such a curve with the smaller one ($i$ = 53 degrees). However, we note that even if we adopt $i$ = 53 degrees, we can get a result that does not affect our conclusion, or the estimates of the scale heights. The other three parameters were adjusted so that $V_{\rm rot}$ and $\sigma_{\rm dis}$ continuously changed across the rings. We determined the free parameters every 0.25 arcsec from the starting point at 0.75 arcsec, given the beam size of $0.52\times0.83$ arcsec$^2$. The innermost part ($\leq 0.5$ arcsec) was not considered because of the weak CO($J$ = 2--1) emission. Our fit took account of channels in the velocity range 1900~km~s$^{-1}$--2800~km~s$^{-1}$ with significance larger than 5$\sigma$ (= 3.8 mJy beam$^{-1}$). 




\begin{figure*}[!t]
  \centering
  \includegraphics[scale=0.3]{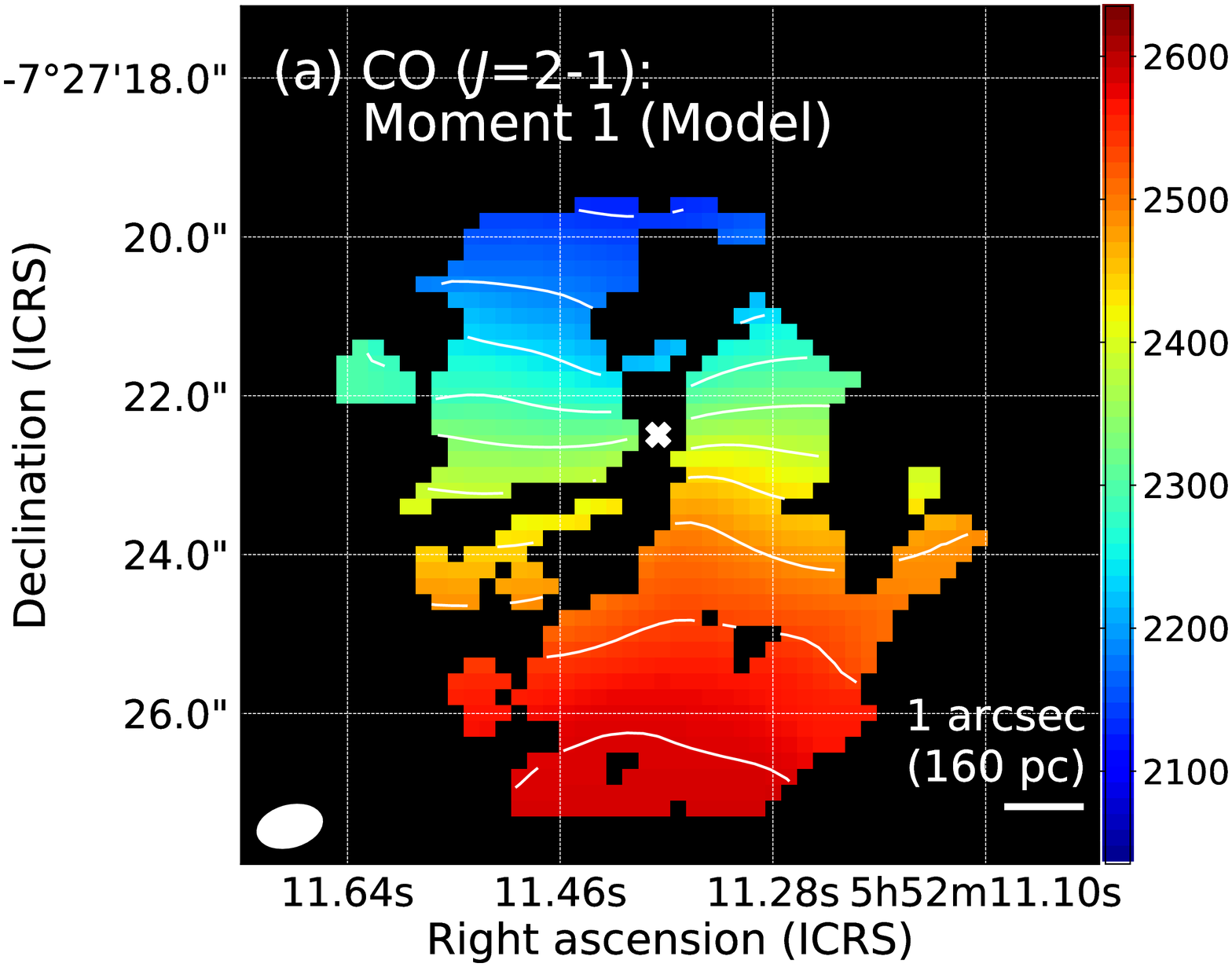}
  \includegraphics[scale=0.3]{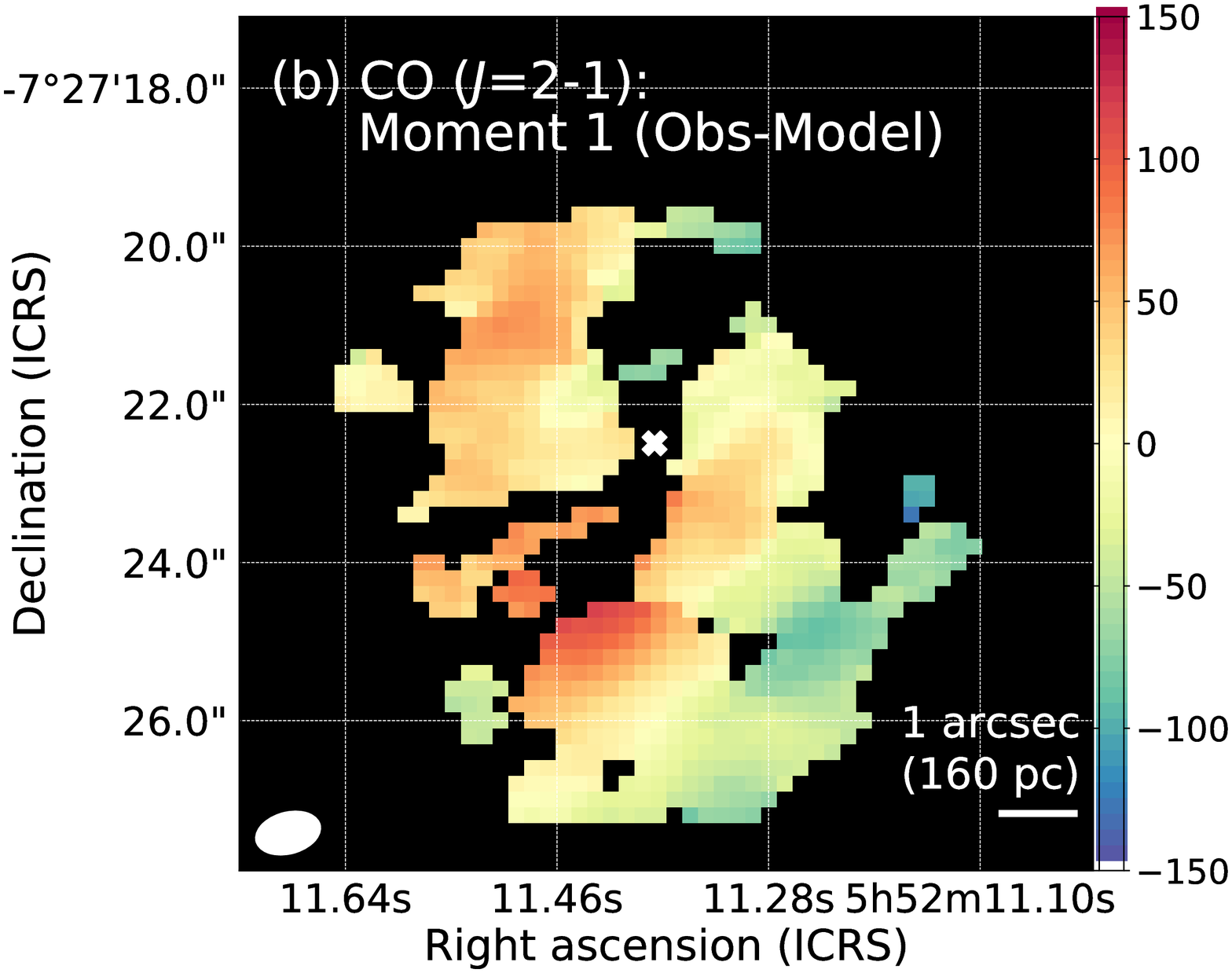}\\
  \caption{\small{
      (a) Model moment 1 map of the CO($J$ = 2--1) emission, where the contours represent the $V_{\rm LSR}$ with steps of 50 km s$^{-1}$ from 2086 km~s$^{-1}$.
      (b) Difference between the observed and modeled moment 1 maps. 
    }
  }\label{fig:ring_model} 
\end{figure*}

\begin{figure}[!t]
  \centering
  \includegraphics[scale=0.32]{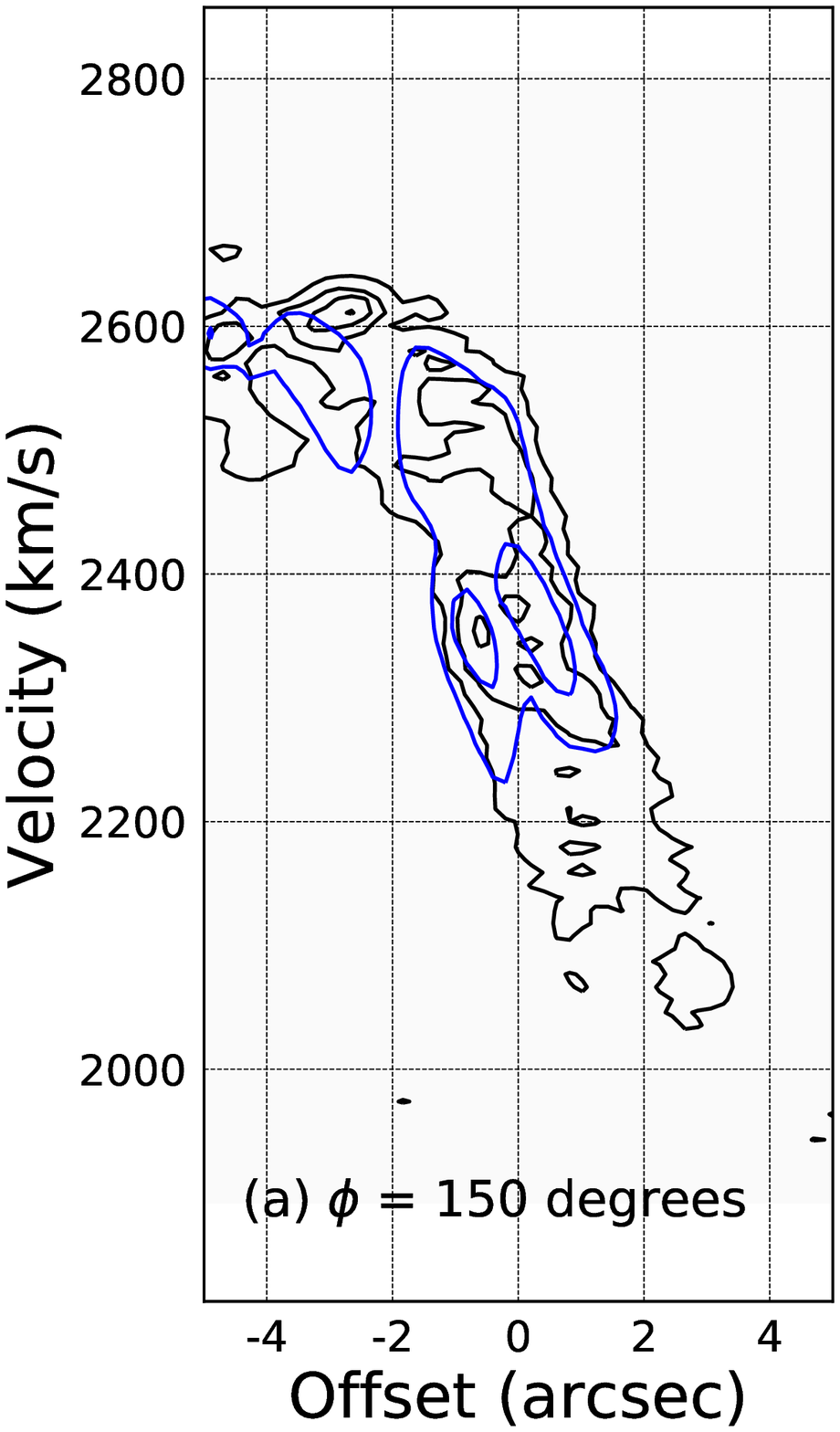}\hspace{-.0cm}
  \includegraphics[scale=0.32]{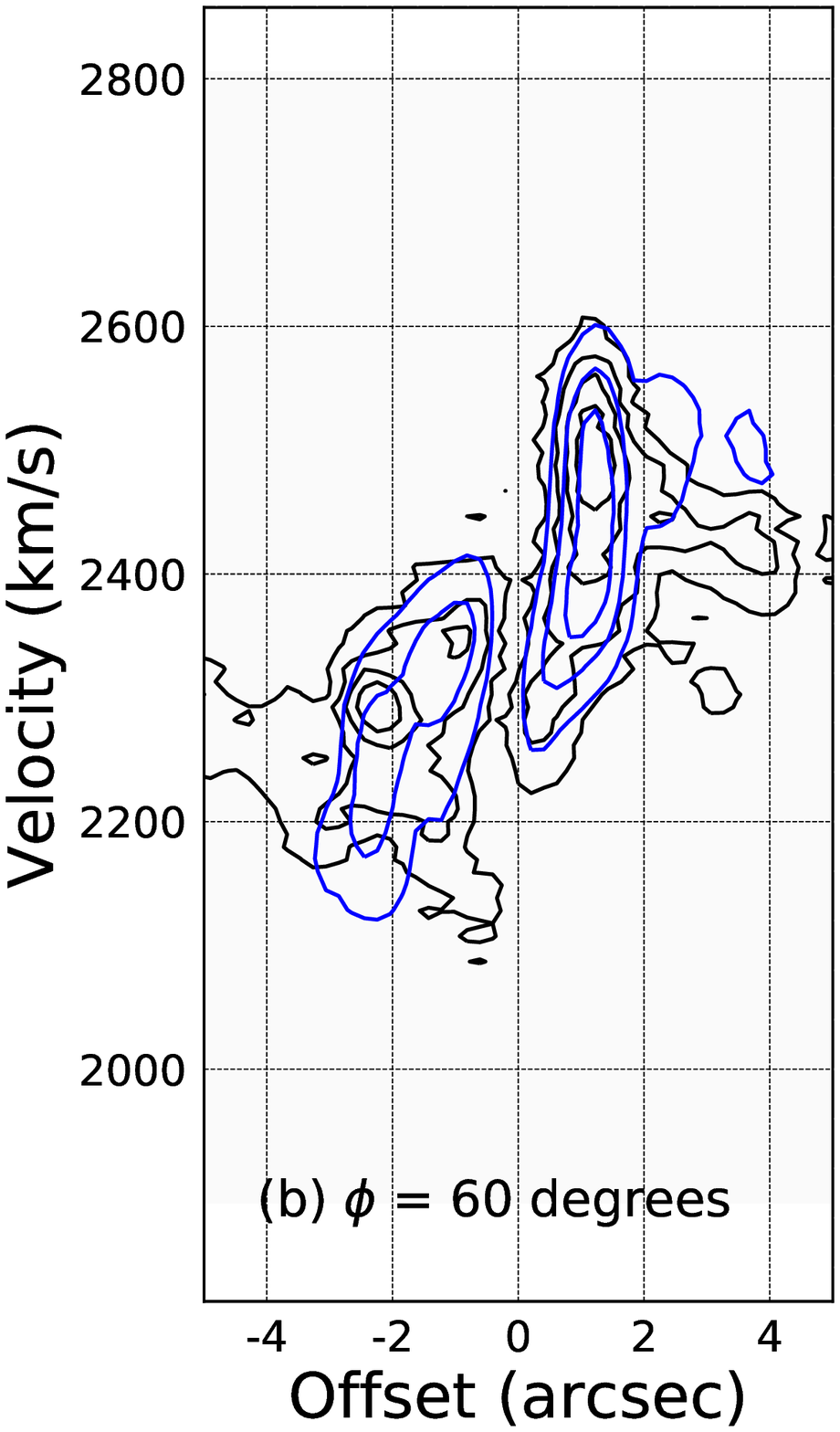}  
  \caption{\small{
      (a) Position-velocity diagrams along position angle of 150 degrees. The model is indicated by the blue contours, while the observed data by the black ones. The contour levels are 1$\sigma$, 3$\sigma$, 5$\sigma$, and 7$\sigma$, where $\sigma$ is 0.76 mJy beam$^{-1}$. A significant ($\sim 7 \sigma$) non-rotating motion with 2600 km s$^{-1}$, or with an excess velocity of 100 km s$^{-1}$, is seen around $-3$ arcsec. 
      (b) Same as (a), but for the position angle of 60 degrees. 
      }
  }\label{fig:pv} 
\end{figure}

\begin{figure}[!t]
  \centering
  \includegraphics[angle=-90,scale=0.4]{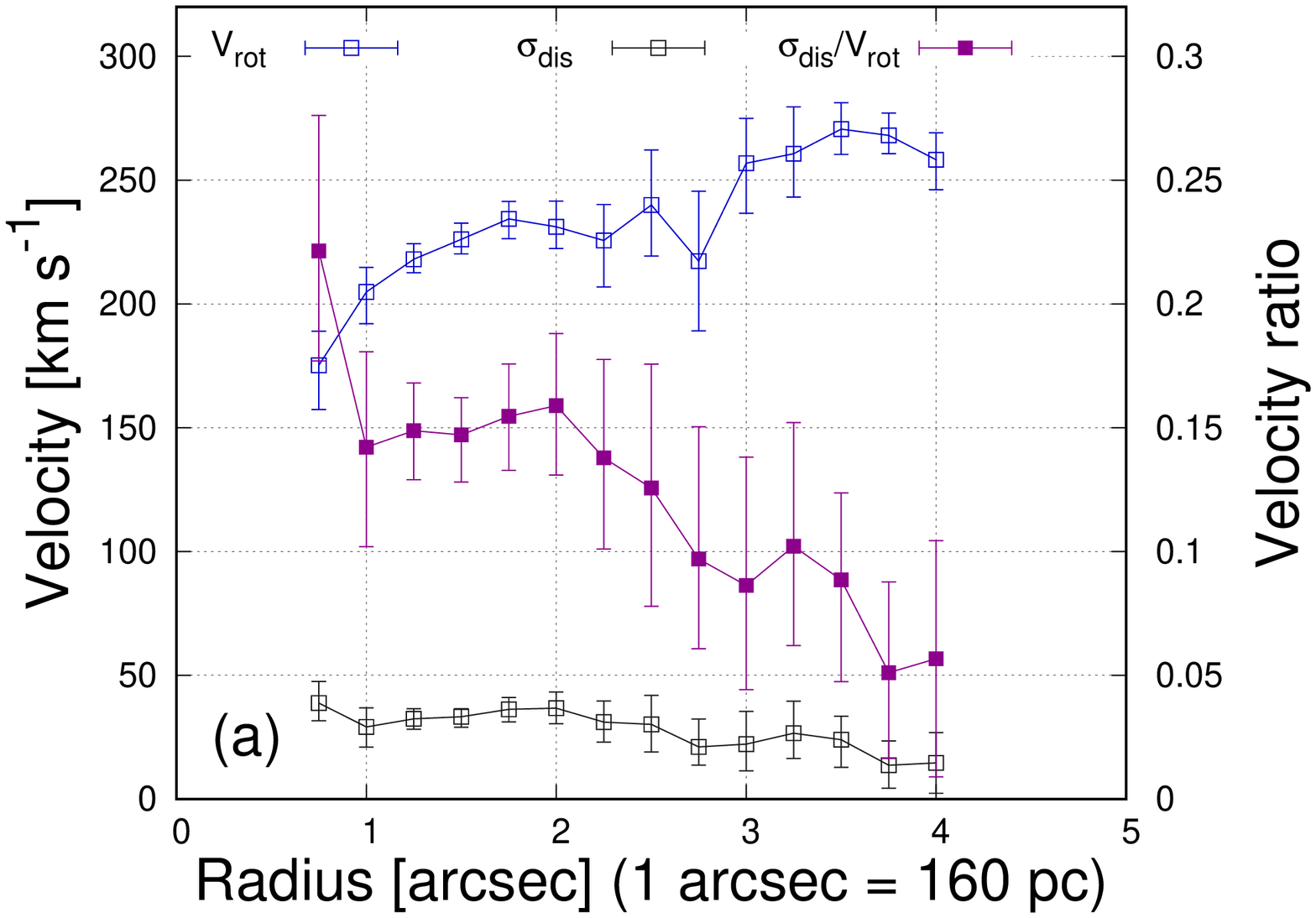}
  \includegraphics[angle=-90,scale=0.38]{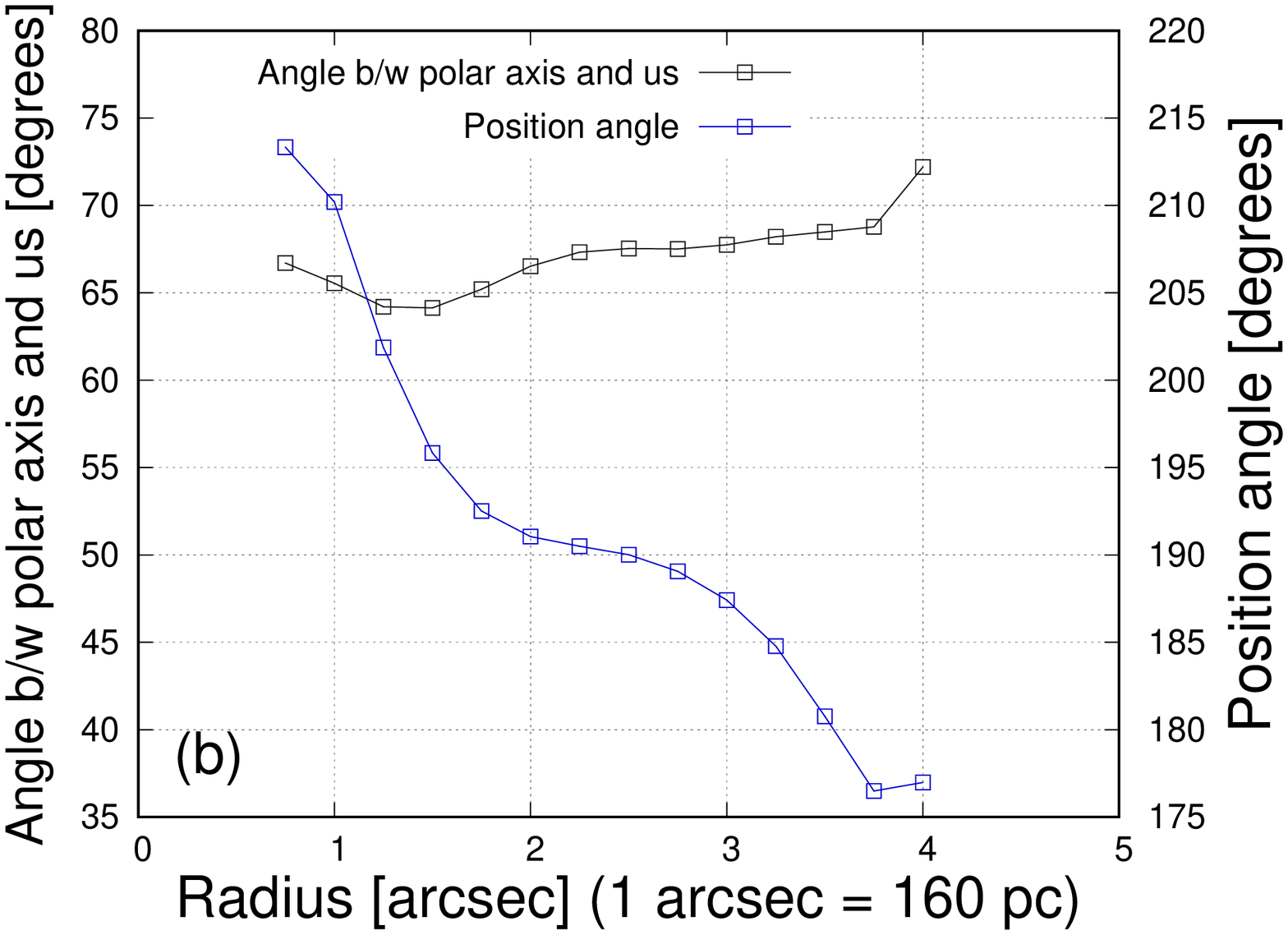}
  \caption{\small{
      (a) Rotation velocity, velocity dispersion, and their ratio as a function of radius from the radio continuum center. 
      (b) Angle between the polar axis and our sightline and position angle as a function of radius from the radio continuum center.
      }
  }\label{fig:ring_pars} 
\end{figure}

The Momonet 1 map of our ring model and residuals against the observed one are seen in Figure~\ref{fig:ring_model}(a) and (b), respectively. Generally, we can see good agreement, except for a southeast red component, as is also seen at $\approx$ 7$\sigma$ in a position velocity diagram of Figure~\ref{fig:pv}(a). Note that the diagrams in Figure~\ref{fig:pv} were created by averaging pixels in slits with 2.5 arcsec width perpendicular to directions of $\phi = 150$ degrees and $\phi = 60$ degrees. The former angle was selected to show the significance of the red component, while the latter was adopted to just show the perpendicular diagram. Interpretation of the red component in excess of $\sim$ 100 km s$^{-1}$ is ambiguous; this may be a high-velocity inflow deviating from an outer rotating disk, or may be an outflow driven by the nuclear radiation. Note that the jet is less likely to be associated with the flow, given its south-north axis (Figure~\ref{fig:cont_map}). The flux density integrated between 2590--2640~km~s$^{-1}$ in a 1.5 arcsec circular region was 1.3 Jy km~s$^{-1}$. By adopting Equation (1) in \cite{Ros19}, the molecular gas mass was estimated to be $\sim10^{6}~M_{\rm sun}$. Given the physical extension of $\sim 3$ arcsec, or 500 pc, the characteristic flow rate was calculated to be $\sim$ 0.2~$M_{\rm sun}$ yr$^{-1}$. Note that the equation we considered was the same as for a thin shell-like geometry (i.e., Equation (4) of \citealt{Mai12}). 

The parameters given by our model are seen as a function of radius from the radio continuum center in Figure~\ref{fig:ring_pars}. The ratios between the rotation velocity and velocity dispersion were $\approx$ 0.05--0.15 in the range 160--640 pc. Therefore, the thicknesses, $\approx2\times$ the scale heights, were estimated to be within $\sim$ 50--110 pc. This range is consistent with the assumed thicknesses (i.e., 80 pc) of the rings.


\section{Discussion}\label{sec:dis}

\begin{figure*}[!t]
  \centering
    \includegraphics[scale=0.60,angle=90]{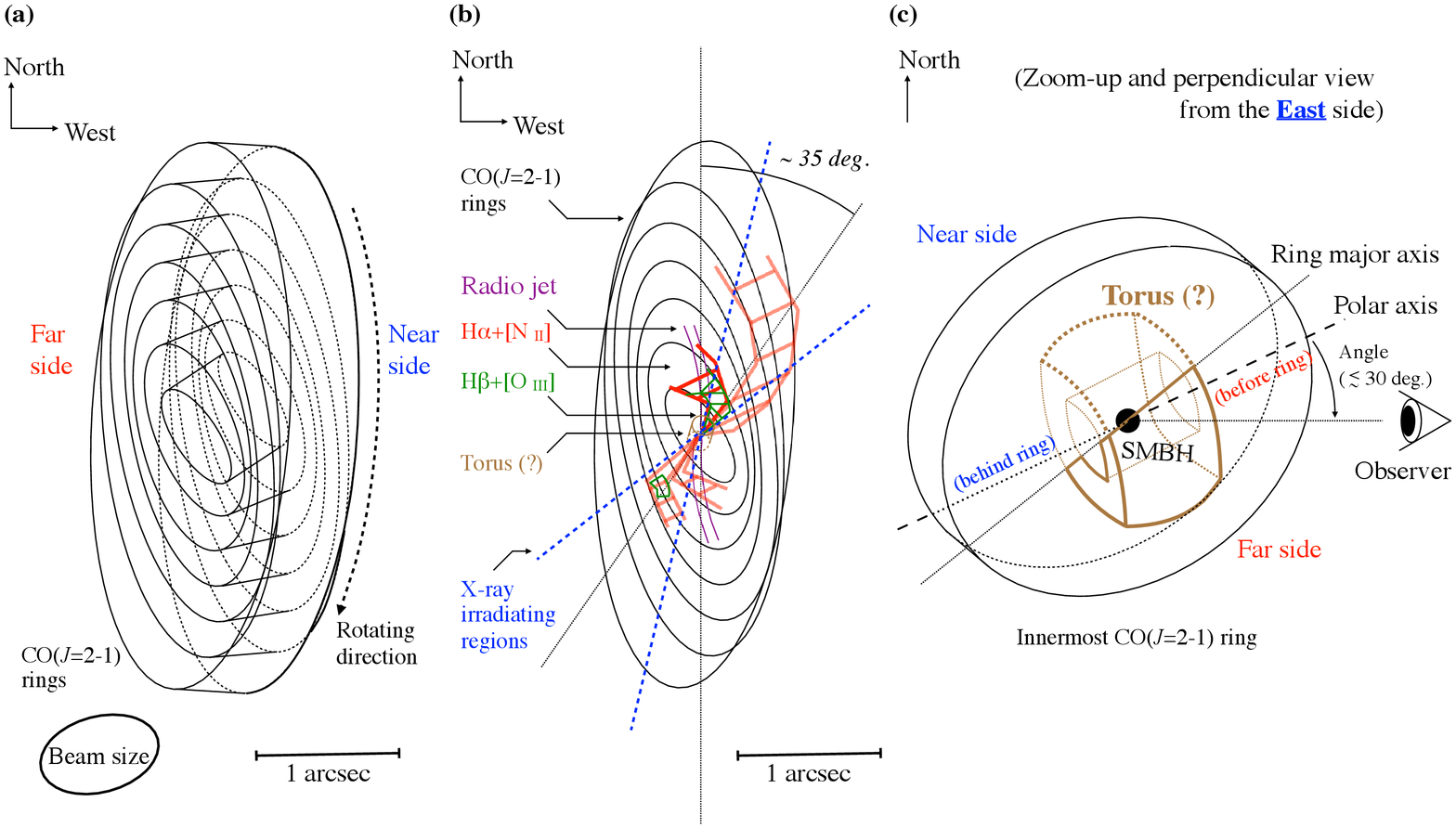}\vspace{-1.3cm}
    \caption{\small{
    (a) Geometrical configuration of the CO($J$ = 2--1) rings 
    in the central $\approx$ 1.3~kpc of NGC~2110. The innermost and outermost rings correspond to those with radii of 0.75 arcsec and 3.75 arcsec, respectively. 
    (b) Spatial comparison between the radio jet \citep[solid purple  lines;][]{Ulv83,Gal99}, the H$\alpha$+[N~II] (red shaded regions) and H$\beta$+[O~III] (green shaded regions) emission \citep{Ros10}, the X-ray irradiating regions (blue dashed lines), and CO($J$ = 2--1) rings (black solid lines). Regarding the H$\alpha$+[N~II] emission, darker red indicates brighter emission. An obscuring torus, speculated from the morphologies of the ionized regions, is also indicated in the center. For clarity, the thicknesses of the CO rings are omitted. 
    (c) Speculated innermost region seen from the East direction. 
    }
  }\label{fig:schematic} 
\end{figure*}

Our discussion is made along with Figure~\ref{fig:schematic}, which shows a detailed geometrical structure of the CO($J$ = 2--1) emission and  our schematic pictures within the central $\approx$ 1.3 kpc of NGC~2110. 

\subsection{Extended Fe-K$\alpha$ emitting region and nuclear obscuration}\label{sec:ext_xray}  

Based on Figure~\ref{fig:x_ima}(b) and (c), we have suggested that the Fe-K$\alpha$ emission is spatially collimated and extends in a northwest-southeast direction out to $\sim$ 3 arcsec ($\sim$ 500 pc) on each side (Figure~\ref{fig:x_ima}). This was also indicated by the spatially-resolved spectral analysis (Figure~\ref{fig:x_spe}). 

To discuss the physical origin, the EW of the Fe-K$\alpha$ emission is useful. As listed in Table~\ref{tab:best_torus}, those in the northwest and southeast regions are as high as $\approx$ 1.5 keV. \modtext{We caution that they do not include the un-resolved nuclear emission and are derived from the extended components.} Such high EWs can be explained by X-ray irradiation of ambient gas. In the case, a direct photo-ionizing X-ray source is not seen in the sightline and only the reflected X-ray emission is seen, thus favoring a high EW \citep[$\sim$ 1~keV; e.g., ][]{Nob10}. It would be fine to rule out another idea that the Fe-K$\alpha$ emission is the result of the scattering of the nuclear emission. This is because the emission line should be accompanied by stronger continuum emission as is seen in un-obscured AGNs \citep[$\sim 200$~eV; ][]{Ric14a,Ric14b}, favoring a lower EW. Also, the collisional ionization would be unlikely as the primary process because a lower EW is expected due to strong bremsstrahlung emission under the line \citep[e.g., ][]{Nob10}. 

The nuclear emission likely plays a major role in the irradiative process, compared with emission from the jet.  Although the beamed X-ray emission from the jet may be more energetic \cite[e.g., ][]{Wea95} and its un-isotropic emission can naturally reproduce a collimated structure, its contribution would be rather minor. The north-south axis of the jet \citep[][]{Ulv83,Nag99} is different from that seen for the Fe-K$\alpha$ emission (Figure~\ref{fig:schematic}). 

Given that the nuclear X-ray emission is more or less  isotropic \citep[e.g., ][]{Liu14}, a collimating structure is needed to be consistent with the observed morphology. A putative AGN torus is a plausible one \citep[e.g., ][]{Urr95}. Its presence was inferred from past optical and NIR observations \citep[e.g., ][]{Ros10,Din15,Din19}. \cite{Ros10} showed that optical emission from ionized atoms (i.e., H$\alpha$+[N~II] and H$\beta$+[O~III]) extends preferentially in a northwest-southeast direction with P.A. $\sim$ $-$35 degrees (Figure~\ref{fig:schematic}(b)). Similar morphologies can be seen at NIR hydrogen emission lines \cite[e.g., Br$\gamma$ and Pa$\beta$; ][]{Din15,Din19}. Thus, as depicted in Figure~\ref{fig:schematic}(b), we suggest that the nuclear X-ray emission is also collimated by the same torus structure as the emission lines in the optical and NIR bands. Note that the X-ray emission may widely irradiate the ambient gas than the longer wavelength emission given the higher penetrating power. 


\begin{figure}[!t]
    \hspace{-.2cm}
    \includegraphics[scale=0.34]{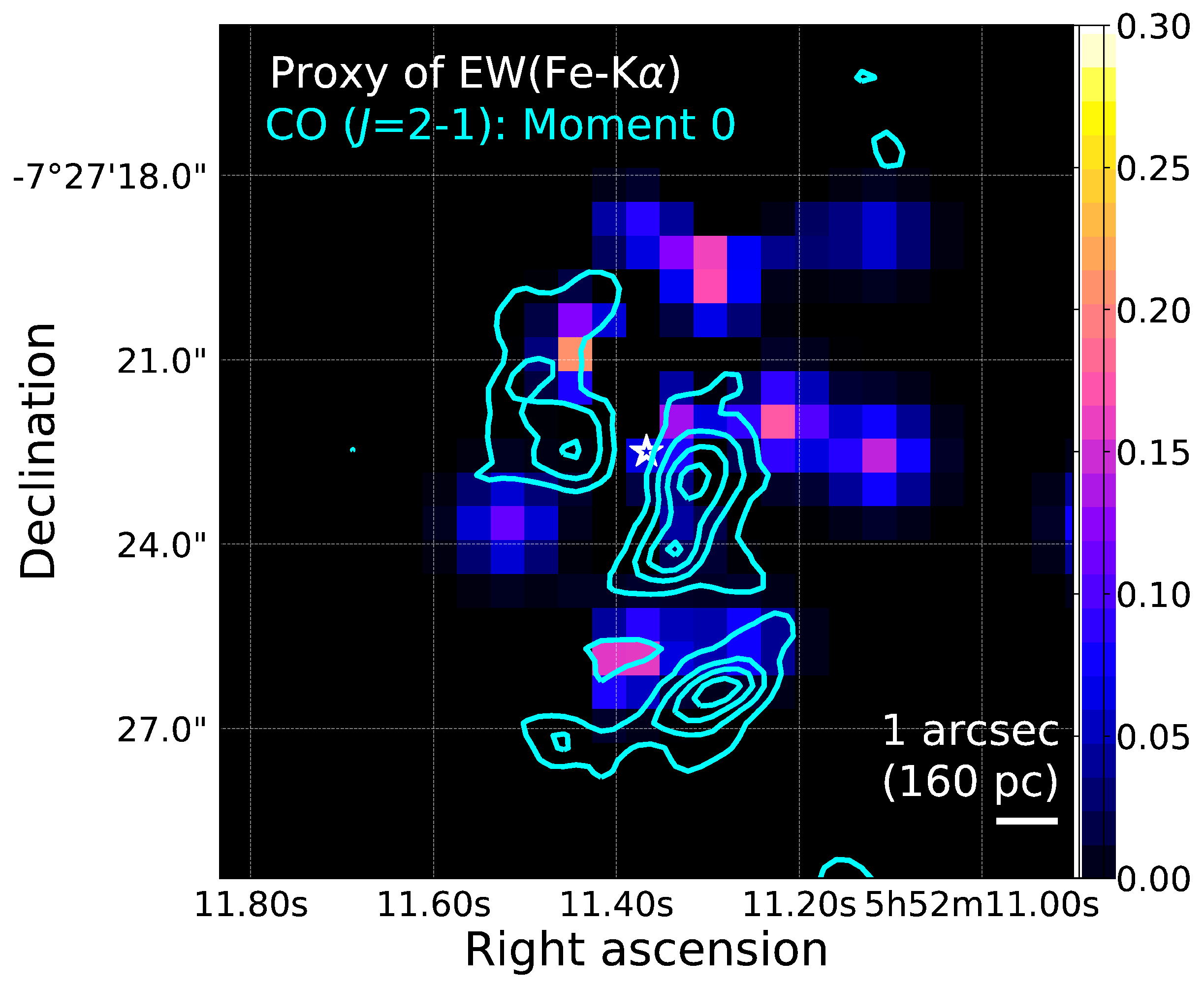}
    \caption{\small{
    Ratio between the 6.2--6.5 keV and 3.0--6.0 keV images (color), a proxy of the Fe-K$\alpha$ line EW, and the CO($J$ = 2--1) velocity-integrated intensity map (cyan), drawn at the same levels in Figure~\ref{fig:mol_map}(a). The white star is located at the 3--7 keV peak. 
    }
  }\label{fig:feco} 
\end{figure}

\subsection{Spatial distributions of Fe-K$\alpha$ and CO($J$ = 2--1) emission and XDR model}\label{sec:xdr} 

The Fe-K$\alpha$ emission is spatially compared with the CO($J$ = 2--1) emission in Figure~\ref{fig:feco}. Their spatial anti-correlation is seen; the CO emission is stronger in the northeast and southwest islands, whereas the Fe-K$\alpha$ emission is bright in the northwest and southeast directions (see Section~\ref{sec:ext_xray}).


We quantitatively discuss the physical origin of the anti-correlation, while referring to an XDR model \citep{Mal96}. The model considered gas illuminated by power-law X-ray emission. The key parameter, which determines fractional abundances of atomic and molecular gas species (see their Figure~3), is the effective ionization parameter, calculated as 
\begin{eqnarray}\label{eqn:ion}
    \xi_{\rm eff} \simeq 0.1 \frac{L_{44}}{n_{4} R^2_{100} N^\alpha_{{\rm H},22}}, 
\end{eqnarray} 
where $L_{44}$, $n_4$, $R_{100}$, $N_{\rm H,22}$, and $\alpha$ represents the 1--100 keV luminosity in units of $10^{44}$ erg s$^{-1}$, the hydrogen molecular gas density in units of $10^{4}$ cm$^{-3}$, the distance the X-ray source in units of 100 pc, the column density of the gas in units of $10^{22}$ cm$^{-2}$ that attenuates the incident X-ray flux, and an X-ray photon index dependent value, respectively. The last factor ($\alpha$) is specifically expressed as $\alpha =$ ($\Gamma$+2/3)/(8/3), taking account of the fact that softer spectra are more heavily extincted. 

Un-absorbed 2--10 keV luminosity was estimated in various epochs, and varied within 0.4--3.5$\times10^{43}$ \ergs \citep{Mar15}. That corresponds to a 1--100 keV luminosity range 2--15$\times10^{43}$ \ergs, where a photon index of 1.65 is assumed. Note that this index corresponds to $\alpha\approx$ 0.87. The density of the molecular hydrogen gas is estimated to be 120--270~cm$^{-3}$ by adopting a hydrogen molecular gas surface density of 650 $M_\odot$ pc$^{-2}$ in the central region with faint CO($J$ = 2--1) emission between the east and west CO islands \citep{Ros19} and the disk thickness ($\approx$ 50--110 pc). The distance $R_{100}$ is 4.8, corresponding to 3 arcsec. Given that distance and the density range, $N_{\rm H}$ $\sim~3\times10^{23}$ cm$^{-2}$ are expected, where we assume the uniform gas density distribution. This is close to the upper bound of those from the spectral fittings in Section~\ref{sec:xspe} (i.e., $N_{\rm H} < 10^{23}$ cm$^{-2}$). As a result, the ionization parameters in NW and SE are conservatively calculated to be $\log \xi_{\rm eff} > -2.4$. 



%


Based on the XDR model by \cite{Mal96}, we can suggest some mechanisms, which suppress CO($J$ = 2--1) emission at the above-estimated ionization parameters. (1) The abundance of CO molecules is decreased due to dissociation by the charge exchange with He ions produced by the direct X-ray ionization. Such dissociation takes place also by Far-UV photons produced as the result of collision between photo-electrons and hydrogen species. (2) Hydrogen molecules are destroyed due to X-ray-produced photo-electrons. Given lower rotational-excitation efficiency of CO by H than by H$_2$ \citep{Gre76}, that would also contribute to the weakness. (3) if achieved, a super-thermal CO population due to additional X-ray heating may also make a contribution. Note that one might consider a low line opacity of CO($J$ = 2--1) as the cause of the weakness. 
This can be achieved if $J$ = 2 level is populated by few CO molecules,
but would be unlikely because $J$ = 2 level can be populated by many CO molecules in energetic circum-nuclear regions due to the low energy gap between $J$ = 2 and $J$ = 1. In summary, via the above three ways, the X-ray emission can weaken the CO($J$ = 2--1) emission. 

To conclude what mechanism is at work, observing other lines is desired. For example, a [C~I] line observation will be useful to examine the CO dissociation scenario by constraining the [C~I]/CO line ratio. To discuss the H$_2$ dissociation, we need to estimate the amount of H$_2$ by a non-CO based method. One way is to utilize rotational quadrupole transition of H$_2$ \citep{Tog16}.
High sensitive spatially-resolved MIR spectroscopy by \textit{JWST}/MIRI will be powerful for the estimate. The last super-thermal scenario will be investigated by observing CO at higher $J$ levels. 




%




We note a discrepancy between an XDR prediction and a suggestion by \cite{Ros19}, who also studied molecular gas in NGC 2110. The XDR model predicts dissociation of an amount of hydrogen molecules into atoms in the NW and SE regions. By contrast,
based on \spitzer/IRS spectroscopy data, \cite{Ros19} suggested the presence of H$_2$ gas with a surface density ($>$ 180 $M_{\rm sun}$ pc$^{-2}$)  comparable to that in the surrounding CO bright regions ($\sim$ 200-350 $M_{\rm sun}$ pc$^{-2}$), disfavoring the dissociation. Further discussion would need  \textit{JWST}/MIRI, if we follow the measurement method of \cite{Ros19}. Briefly, they first determined a power-law temperature distribution consistent with observed H$_2$ populations at different excitation levels. They then derived the surface density while taking account of H$_2$ populations down to 50~K. This derivation is sensitive to the intensity at the lowest permitted rotational (quadrupole) transition of H$_2$ 0--0 S(0). However, only an upper-limit was obtained for the transition. Thus, its exact intensity constrained by MIRI is desired for further discussion.


\subsection{Geometrical structures inferred from comparison between CO($J$ = 2--1) emission and optical one}\label{sec:co_optir} 

We note our serendipitous finding in Figure~\ref{fig:schematic} that the distribution of the H$\alpha$+[N~II] emission, drawn based on Figure~3 in \cite{Ros10}, looks like following some of the CO($J$ = 2--1) tilted rings obtained in Section~\ref{sec:kin}. This may be a natural consequence of the irradiation by the nuclear source. Given the canonical definition of the ionization parameter (i.e., $\propto L/nR^2$, where $L$, $n$, and $R$ are the luminosity, density, and distance, respectively), an ionized region can extend up to a distance under a uniform gas distribution, consistent with the observed. Such was already argued in the past \citep[e.g., ][]{Din15}. However, we emphasize that our constraint on the three-dimensional structure of the CO gas makes that picture clearer. 



Also, we remark that from the configuration of the CO gas and ionized gas, it is possible to estimate an angle between our sightline and the polar axis of the ionization cone, or that of the torus. \cite{Ros10} suggested that the southern [O~III] emission fainter than the northern one was due to dust extinction. Under the assumption that the dust distributes similarly to the molecular gas, it is suggested that the southern (the northern) polar axis of the ionization cone is preferentially located behind (before) the southern (northern) gas rings (Figure~\ref{fig:schematic}(c)). To  accomplish such a configuration, the angle is required to be $\lesssim~30$ degrees. The AGN torus may have a half-opening angle of $\approx~37$ degrees, as estimated from its Eddington ratio \citep[$\log \lambda_{\rm Edd} \sim -2$; ][]{Kaw16b} and a correlation between the Eddington ratio and the half-opening angle \citep{Ric17nat}. These results suggest that we see the nucleus almost at an intermediate angle between the obscuring torus and a region with less obscuration. Because of this angle, we may have often observed column-density variability and various obscuring components in the X-ray band \citep{Riv14}. 



\section{Summary}\label{sec:sum}

We have analyzed high-angular resolution ($\sim$ 0.5 arcsec) \chandra and ALMA data of an obscured AGN host of NGC 2110 to investigate AGN X-ray irradiation effects on the surrounding ISM. Our findings are summarized as follows. 

\begin{itemize}
    \item An $\sim$ 7\% of the Fe-K$\alpha$ emission was spatially resolved by \chandra and extends preferentially in a northwest-to-southeast direction out to $\sim$ 3 arcsec (Figure~\ref{fig:x_ima}(c) and Section~\ref{sec:ext_xray}). The EWs of the Fe-K$\alpha$ emission in the NW and SE regions were $\sim$ 1.5 keV (Figure~\ref{fig:x_spe} and Table~\ref{tab:best_pow}), high enough to indicate X-ray irradiation as the physical origin. 
    
    \item The tentative spatial anti-correlation between the Fe-K$\alpha$ and CO($J$ = 2--1) emission was found (Figure~\ref{fig:feco}). 
    
    \item The anti-correlation was discussed based on the ionization parameter defined in the XDR model (Section~\ref{sec:xdr}). The derived parameters predict some mechanisms at work that weaken the CO($J$ = 2--1) emission. The CO and/or H$_2$ molecules may be dissociated due to X-ray emission. Also, CO molecules may be super-thermalized by X-ray emission, making the low rotation level $J$ = 2--1 line weak. Follow-up observations of other lines (e.g., [C~I], H$_2$ 0-0 S(0), and higher-$J$ CO lines) will be helpful for further discussion. As a final goal, such studies would lead to discussion on whether the X-ray irradiation eventually affects the surrounding star formation as an AGN feedback.
    
    \item Based on the geometrical structures of the molecular gas disk and the ionized regions, it was inferred that the polar axis of the ionization field, or the torus, is inclined by an angle $\lesssim 30$ degrees (Section~\ref{sec:co_optir} and Figure~\ref{fig:schematic}(c)). 
\end{itemize}

In the future, sub-arcsec resolution observatories with larger effective areas (i.e., \textit{AXIS} and \textit{Lynx}; \citealt{Ray17} and \citealt{Gas18}) will be launched. In such an era, they will enable us to more clearly map the Fe-K$\alpha$ line distribution of NGC~2110. Also, it will be possible to discuss XDRs in distant, more luminous AGNs, which have potential to more largely affect the surrounding ISM. 







\acknowledgments

Part of this work was financially supported by the Grant-in-Aid for JSPS Fellows for young researchers (T.K. and S.B.). T.K., T.I. and M.I. are supported by JSPS KAKENHI grant numbers 17J09016, 17K14247, and 15K05030, respectively. The scientific results reported in this article are based on data obtained from the Chandra Data Archive. This research has made use of software provided by the Chandra X-ray Center (CXC) in the application packages CIAO. This paper makes use of the following ALMA data: ADS/JAO.ALMA\#2012.1.00474. ALMA is a partnership of ESO (representing its member states), NSF (USA) and NINS (Japan), together with NRC (Canada) and NSC and ASIAA (Taiwan) and KASI (Republic of Korea), in cooperation with the Republic of Chile. The Joint ALMA Observatory is operated by ESO, AUI/NRAO and NAOJ. We appreciate the JVO portal (http://jvo.nao.ac.jp/portal/) operated by ADC/NAOJ for the quick look of the ALMA archive data. 

\bibliographystyle{aasjournal}
\bibliography{ref}

\begin{thebibliography}{}
\expandafter\ifx\csname natexlab\endcsname\relax\def\natexlab#1{#1}\fi
\providecommand{\url}[1]{\href{#1}{#1}}
\providecommand{\dodoi}[1]{doi:~\href{http://doi.org/#1}{\nolinkurl{#1}}}
\providecommand{\doeprint}[1]{\href{http://ascl.net/#1}{\nolinkurl{http://ascl.net/#1}}}
\providecommand{\doarXiv}[1]{\href{https://arxiv.org/abs/#1}{\nolinkurl{https://arxiv.org/abs/#1}}}

\bibitem[{{Bigiel} {et~al.}(2008){Bigiel}, {Leroy}, {Walter}, {Brinks}, {de
  Blok}, {Madore}, \& {Thornley}}]{Big08}
{Bigiel}, F., {Leroy}, A., {Walter}, F., {et~al.} 2008, \aj, 136, 2846,
  \dodoi{10.1088/0004-6256/136/6/2846}

\bibitem[{{Bradt} {et~al.}(1978){Bradt}, {Burke}, {Canizares}, {Greenfield},
  {Kelley}, {McClintock}, {van Paradijs}, \& {Koski}}]{Bra78}
{Bradt}, H.~V., {Burke}, B.~F., {Canizares}, C.~R., {et~al.} 1978, \apjl, 226,
  L111, \dodoi{10.1086/182843}

\bibitem[{{Byrne} {et~al.}(2019){Byrne}, {Christensen}, {Tsekitsidis},
  {Brooks}, \& {Quinn}}]{Byr19}
{Byrne}, L., {Christensen}, C., {Tsekitsidis}, M., {Brooks}, A., \& {Quinn}, T.
  2019, \apj, 871, 213, \dodoi{10.3847/1538-4357/aaf9aa}

\bibitem[{{Canizares} {et~al.}(2005){Canizares}, {Davis}, {Dewey}, {Flanagan},
  {Galton}, {Huenemoerder}, {Ishibashi}, {Markert}, {Marshall}, {McGuirk},
  {Schattenburg}, {Schulz}, {Smith}, \& {Wise}}]{Can05}
{Canizares}, C.~R., {Davis}, J.~E., {Dewey}, D., {et~al.} 2005, \pasp, 117,
  1144, \dodoi{10.1086/432898}

\bibitem[{{Cash}(1979)}]{Cas79}
{Cash}, W. 1979, \apj, 228, 939, \dodoi{10.1086/156922}

\bibitem[{{Clements}(1983)}]{Cle83}
{Clements}, E.~D. 1983, \mnras, 204, 811, \dodoi{10.1093/mnras/204.3.811}

\bibitem[{{Davis} {et~al.}(2012){Davis}, {Bautz}, {Dewey}, {Heilmann}, {Houck},
  {Huenemoerder}, {Marshall}, {Nowak}, {Schattenburg}, {Schulz}, \&
  {Smith}}]{Dav12}
{Davis}, J.~E., {Bautz}, M.~W., {Dewey}, D., {et~al.} 2012, in Society of
  Photo-Optical Instrumentation Engineers (SPIE) Conference Series, Vol. 8443,
  \procspie, 84431A

\bibitem[{{de Vaucouleurs} {et~al.}(1991){de Vaucouleurs}, {de Vaucouleurs},
  {Corwin}, {Buta}, {Paturel}, \& {Fouque}}]{Dev91}
{de Vaucouleurs}, G., {de Vaucouleurs}, A., {Corwin}, Herold~G., J., {et~al.}
  1991, {Third Reference Catalogue of Bright Galaxies}

\bibitem[{{Di Teodoro} \& {Fraternali}(2015)}]{Dit15}
{Di Teodoro}, E.~M., \& {Fraternali}, F. 2015, \mnras, 451, 3021,
  \dodoi{10.1093/mnras/stv1213}

\bibitem[{{Diniz} {et~al.}(2019){Diniz}, {Riffel}, {Storchi-Bergmann}, \&
  {Riffel}}]{Din19}
{Diniz}, M.~R., {Riffel}, R.~A., {Storchi-Bergmann}, T., \& {Riffel}, R. 2019,
  \mnras, 487, 3958, \dodoi{10.1093/mnras/stz1329}

\bibitem[{{Diniz} {et~al.}(2015){Diniz}, {Riffel}, {Storchi-Bergmann}, \&
  {Winge}}]{Din15}
{Diniz}, M.~R., {Riffel}, R.~A., {Storchi-Bergmann}, T., \& {Winge}, C. 2015,
  \mnras, 453, 1727, \dodoi{10.1093/mnras/stv1694}

\bibitem[{{Durr{\'e}} \& {Mould}(2014)}]{Dur14}
{Durr{\'e}}, M., \& {Mould}, J. 2014, \apj, 784, 79,
  \dodoi{10.1088/0004-637X/784/1/79}

\bibitem[{{Evans} {et~al.}(2006){Evans}, {Lee}, {Kamenetska}, {Gallagher},
  {Kraft}, {Hardcastle}, \& {Weaver}}]{Eva06}
{Evans}, D.~A., {Lee}, J.~C., {Kamenetska}, M., {et~al.} 2006, \apj, 653, 1121,
  \dodoi{10.1086/508680}

\bibitem[{{Evans} {et~al.}(2007){Evans}, {Lee}, {Turner}, {Weaver}, \&
  {Marshall}}]{Eva07}
{Evans}, D.~A., {Lee}, J.~C., {Turner}, T.~J., {Weaver}, K.~A., \& {Marshall},
  H.~L. 2007, \apj, 671, 1345, \dodoi{10.1086/523037}

\bibitem[{{Fabbiano} {et~al.}(2017){Fabbiano}, {Elvis}, {Paggi}, {Karovska},
  {Maksym}, {Raymond}, {Risaliti}, \& {Wang}}]{Fab17}
{Fabbiano}, G., {Elvis}, M., {Paggi}, A., {et~al.} 2017, \apjl, 842, L4,
  \dodoi{10.3847/2041-8213/aa7551}

\bibitem[{{Fabbiano} {et~al.}(2019{\natexlab{a}}){Fabbiano}, {Paggi}, \&
  {Elvis}}]{Fab19b}
{Fabbiano}, G., {Paggi}, A., \& {Elvis}, M. 2019{\natexlab{a}}, \apjl, 876,
  L18, \dodoi{10.3847/2041-8213/ab1c63}

\bibitem[{{Fabbiano} {et~al.}(2019{\natexlab{b}}){Fabbiano}, {Siemiginowska},
  {Paggi}, {Elvis}, {Volonteri}, {Mayer}, {Karovska}, {Maksym}, {Risaliti}, \&
  {Wang}}]{Fab19a}
{Fabbiano}, G., {Siemiginowska}, A., {Paggi}, A., {et~al.} 2019{\natexlab{b}},
  \apj, 870, 69, \dodoi{10.3847/1538-4357/aaf0a4}

\bibitem[{{Ferrarese} \& {Merritt}(2000)}]{Fer00}
{Ferrarese}, L., \& {Merritt}, D. 2000, \apjl, 539, L9, \dodoi{10.1086/312838}

\bibitem[{{Ferruit} {et~al.}(2004){Ferruit}, {Mundell}, {Nagar}, {Emsellem},
  {P{\'e}contal}, {Wilson}, \& {Schinnerer}}]{Fer04}
{Ferruit}, P., {Mundell}, C.~G., {Nagar}, N.~M., {et~al.} 2004, \mnras, 352,
  1180, \dodoi{10.1111/j.1365-2966.2004.08009.x}

\bibitem[{{Gallimore} {et~al.}(1999){Gallimore}, {Baum}, {O'Dea}, {Pedlar}, \&
  {Brinks}}]{Gal99}
{Gallimore}, J.~F., {Baum}, S.~A., {O'Dea}, C.~P., {Pedlar}, A., \& {Brinks},
  E. 1999, \apj, 524, 684, \dodoi{10.1086/307853}

\bibitem[{{Garc{\'\i}a-Burillo} {et~al.}(2010){Garc{\'\i}a-Burillo}, {Usero},
  {Fuente}, {Mart{\'\i}n-Pintado}, {Boone}, {Aalto}, {Krips}, {Neri},
  {Schinnerer}, \& {Tacconi}}]{Gal10}
{Garc{\'\i}a-Burillo}, S., {Usero}, A., {Fuente}, A., {et~al.} 2010, \aap, 519,
  A2, \dodoi{10.1051/0004-6361/201014539}

\bibitem[{{Garmire} {et~al.}(2003){Garmire}, {Bautz}, {Ford}, {Nousek}, \&
  {Ricker}}]{Gar03}
{Garmire}, G.~P., {Bautz}, M.~W., {Ford}, P.~G., {Nousek}, J.~A., \& {Ricker},
  George~R., J. 2003, in Society of Photo-Optical Instrumentation Engineers
  (SPIE) Conference Series, Vol. 4851, \procspie, ed. J.~E. {Truemper} \& H.~D.
  {Tananbaum}, 28--44

\bibitem[{{Gaskin} {et~al.}(2018){Gaskin}, {Dominguez}, {Gelmis}, {Mulqueen},
  {Swartz}, {McCarley}, {{\"O}zel}, {Vikhlinin}, {Schwartz}, {Tananbaum},
  {Blackwood}, {Arenberg}, {Purcell}, \& {Allen}}]{Gas18}
{Gaskin}, J.~A., {Dominguez}, A., {Gelmis}, K., {et~al.} 2018, in Society of
  Photo-Optical Instrumentation Engineers (SPIE) Conference Series, Vol. 10699,
  \procspie, 106990N

\bibitem[{{Gebhardt} {et~al.}(2000){Gebhardt}, {Bender}, {Bower}, {Dressler},
  {Faber}, {Filippenko}, {Green}, {Grillmair}, {Ho}, {Kormendy}, {Lauer},
  {Magorrian}, {Pinkney}, {Richstone}, \& {Tremaine}}]{Geb00}
{Gebhardt}, K., {Bender}, R., {Bower}, G., {et~al.} 2000, \apjl, 539, L13,
  \dodoi{10.1086/312840}

\bibitem[{{Gehrels}(1986)}]{Geh86}
{Gehrels}, N. 1986, \apj, 303, 336, \dodoi{10.1086/164079}

\bibitem[{{Glover} \& {Clark}(2012)}]{Glo12}
{Glover}, S. C.~O., \& {Clark}, P.~C. 2012, \mnras, 421, 9,
  \dodoi{10.1111/j.1365-2966.2011.19648.x}

\bibitem[{{G{\'o}mez-Guijarro} {et~al.}(2017){G{\'o}mez-Guijarro},
  {Gonz{\'a}lez-Mart{\'\i}n}, {Ramos Almeida}, {Rodr{\'\i}guez-Espinosa}, \&
  {Gallego}}]{Gom17}
{G{\'o}mez-Guijarro}, C., {Gonz{\'a}lez-Mart{\'\i}n}, O., {Ramos Almeida}, C.,
  {Rodr{\'\i}guez-Espinosa}, J.~M., \& {Gallego}, J. 2017, \mnras, 469, 2720,
  \dodoi{10.1093/mnras/stx1037}

\bibitem[{{Gonz{\'a}lez Delgado} {et~al.}(2002){Gonz{\'a}lez Delgado},
  {Arribas}, {P{\'e}rez}, \& {Heckman}}]{Del02}
{Gonz{\'a}lez Delgado}, R.~M., {Arribas}, S., {P{\'e}rez}, E., \& {Heckman}, T.
  2002, \apj, 579, 188, \dodoi{10.1086/342675}

\bibitem[{{Green} \& {Thaddeus}(1976)}]{Gre76}
{Green}, S., \& {Thaddeus}, P. 1976, \apj, 205, 766, \dodoi{10.1086/154333}

\bibitem[{{Hayashi} {et~al.}(1996){Hayashi}, {Koyama}, {Awaki}, \&
  {Yamauchi}}]{Hay96}
{Hayashi}, I., {Koyama}, K., {Awaki}, H., \& {Yamauchi}, S. U.~S. 1996, \pasj,
  48, 219, \dodoi{10.1093/pasj/48.2.219}

\bibitem[{{Hocuk} \& {Spaans}(2010)}]{Hoc10}
{Hocuk}, S., \& {Spaans}, M. 2010, \aap, 522, A24,
  \dodoi{10.1051/0004-6361/201015055}

\bibitem[{{Hocuk} \& {Spaans}(2011)}]{Hoc11}
---. 2011, \aap, 536, A41, \dodoi{10.1051/0004-6361/201117431}

\bibitem[{{Ikeda} {et~al.}(2009){Ikeda}, {Awaki}, \& {Terashima}}]{Ike09}
{Ikeda}, S., {Awaki}, H., \& {Terashima}, Y. 2009, \apj, 692, 608,
  \dodoi{10.1088/0004-637X/692/1/608}

\bibitem[{{Izumi} {et~al.}(2015){Izumi}, {Kohno}, {Aalto}, {Doi}, {Espada},
  {Fathi}, {Harada}, {Hatsukade}, {Hattori}, {Hsieh}, {Ikarashi}, {Imanishi},
  {Iono}, {Ishizuki}, {Krips}, {Mart{\'\i}n}, {Matsushita}, {Meier}, {Nagai},
  {Nakai}, {Nakajima}, {Nakanishi}, {Nomura}, {Regan}, {Schinnerer}, {Sheth},
  {Takano}, {Tamura}, {Terashima}, {Tosaki}, {Turner}, {Umehata}, \&
  {Wiklind}}]{Izu15}
{Izumi}, T., {Kohno}, K., {Aalto}, S., {et~al.} 2015, \apj, 811, 39,
  \dodoi{10.1088/0004-637X/811/1/39}

\bibitem[{{Kaastra}(2017)}]{Kaa17}
{Kaastra}, J.~S. 2017, \aap, 605, A51, \dodoi{10.1051/0004-6361/201629319}

\bibitem[{{Kawamuro} {et~al.}(2019){Kawamuro}, {Izumi}, \& {Imanishi}}]{Kaw19b}
{Kawamuro}, T., {Izumi}, T., \& {Imanishi}, M. 2019, \pasj, 71, 68,
  \dodoi{10.1093/pasj/psz045}

\bibitem[{{Kawamuro} {et~al.}(2016){Kawamuro}, {Ueda}, {Tazaki}, {Ricci}, \&
  {Terashima}}]{Kaw16b}
{Kawamuro}, T., {Ueda}, Y., {Tazaki}, F., {Ricci}, C., \& {Terashima}, Y. 2016,
  \apjs, 225, 14, \dodoi{10.3847/0067-0049/225/1/14}

\bibitem[{{Kawamuro} {et~al.}(2013){Kawamuro}, {Ueda}, {Tazaki}, \&
  {Terashima}}]{Kaw13}
{Kawamuro}, T., {Ueda}, Y., {Tazaki}, F., \& {Terashima}, Y. 2013, \apj, 770,
  157, \dodoi{10.1088/0004-637X/770/2/157}

\bibitem[{{Kawamuro} {et~al.}(2018){Kawamuro}, {Ueda}, {Shidatsu}, {Hori},
  {Morii}, {Nakahira}, {Isobe}, {Kawai}, {Mihara}, {Matsuoka}, {Morita},
  {Nakajima}, {Negoro}, {Oda}, {Sakamoto}, {Serino}, {Sugizaki}, {Tanimoto},
  {Tomida}, {Tsuboi}, {Tsunemi}, {Ueno}, {Yamaoka}, {Yamada}, {Yoshida},
  {Iwakiri}, {Kawakubo}, {Sugawara}, {Sugita}, {Tachibana}, \&
  {Yoshii}}]{Kaw18}
{Kawamuro}, T., {Ueda}, Y., {Shidatsu}, M., {et~al.} 2018, \apjs, 238, 32,
  \dodoi{10.3847/1538-4365/aad1ef}

\bibitem[{{Kennicutt} {et~al.}(2007){Kennicutt}, {Calzetti}, {Walter}, {Helou},
  {Hollenbach}, {Armus}, {Bendo}, {Dale}, {Draine}, {Engelbracht}, {Gordon},
  {Prescott}, {Regan}, {Thornley}, {Bot}, {Brinks}, {de Blok}, {de Mello},
  {Meyer}, {Moustakas}, {Murphy}, {Sheth}, \& {Smith}}]{Ken07}
{Kennicutt}, Robert~C., J., {Calzetti}, D., {Walter}, F., {et~al.} 2007, \apj,
  671, 333, \dodoi{10.1086/522300}

\bibitem[{{Kormendy} \& {Ho}(2013)}]{Kor13}
{Kormendy}, J., \& {Ho}, L.~C. 2013, \araa, 51, 511,
  \dodoi{10.1146/annurev-astro-082708-101811}

\bibitem[{{Koss} {et~al.}(2011){Koss}, {Mushotzky}, {Veilleux}, {Winter},
  {Baumgartner}, {Tueller}, {Gehrels}, \& {Valencic}}]{Kos11}
{Koss}, M., {Mushotzky}, R., {Veilleux}, S., {et~al.} 2011, \apj, 739, 57,
  \dodoi{10.1088/0004-637X/739/2/57}

\bibitem[{{Krolik} \& {Kallman}(1983)}]{Kro83}
{Krolik}, J.~H., \& {Kallman}, T.~R. 1983, \apj, 267, 610,
  \dodoi{10.1086/160897}

\bibitem[{{LaMassa} {et~al.}(2012){LaMassa}, {Heckman}, \& {Ptak}}]{LaM12}
{LaMassa}, S.~M., {Heckman}, T.~M., \& {Ptak}, A. 2012, \apj, 758, 82,
  \dodoi{10.1088/0004-637X/758/2/82}

\bibitem[{{LaMassa} {et~al.}(2017){LaMassa}, {Yaqoob}, {Levenson}, {Boorman},
  {Heckman}, {Gandhi}, {Rigby}, {Urry}, \& {Ptak}}]{Lam17}
{LaMassa}, S.~M., {Yaqoob}, T., {Levenson}, N.~A., {et~al.} 2017, \apj, 835,
  91, \dodoi{10.3847/1538-4357/835/1/91}

\bibitem[{{Lepp} \& {Dalgarno}(1996)}]{Lep96}
{Lepp}, S., \& {Dalgarno}, A. 1996, \aap, 306, L21

\bibitem[{{Li} {et~al.}(2003){Li}, {Kastner}, {Prigozhin}, \& {Schulz}}]{Li03}
{Li}, J., {Kastner}, J.~H., {Prigozhin}, G.~Y., \& {Schulz}, N.~S. 2003, \apj,
  590, 586, \dodoi{10.1086/374967}

\bibitem[{{Li} {et~al.}(2004){Li}, {Kastner}, {Prigozhin}, {Schulz},
  {Feigelson}, \& {Getman}}]{Li04}
{Li}, J., {Kastner}, J.~H., {Prigozhin}, G.~Y., {et~al.} 2004, \apj, 610, 1204,
  \dodoi{10.1086/421866}

\bibitem[{{Liu} {et~al.}(2014){Liu}, {Wang}, {Yang}, {Zhu}, \& {Zhou}}]{Liu14}
{Liu}, T., {Wang}, J.-X., {Yang}, H., {Zhu}, F.-F., \& {Zhou}, Y.-Y. 2014,
  \apj, 783, 106, \dodoi{10.1088/0004-637X/783/2/106}

\bibitem[{{Magorrian} {et~al.}(1998){Magorrian}, {Tremaine}, {Richstone},
  {Bender}, {Bower}, {Dressler}, {Faber}, {Gebhardt}, {Green}, {Grillmair},
  {Kormendy}, \& {Lauer}}]{Mag98}
{Magorrian}, J., {Tremaine}, S., {Richstone}, D., {et~al.} 1998, \aj, 115,
  2285, \dodoi{10.1086/300353}

\bibitem[{{Maiolino} {et~al.}(2012){Maiolino}, {Gallerani}, {Neri}, {Cicone},
  {Ferrara}, {Genzel}, {Lutz}, {Sturm}, {Tacconi}, {Walter}, {Feruglio},
  {Fiore}, \& {Piconcelli}}]{Mai12}
{Maiolino}, R., {Gallerani}, S., {Neri}, R., {et~al.} 2012, \mnras, 425, L66,
  \dodoi{10.1111/j.1745-3933.2012.01303.x}

\bibitem[{{Maloney}(1999)}]{Mal99}
{Maloney}, P.~R. 1999, \apss, 266, 207.
\newblock \doarXiv{astro-ph/9903275}

\bibitem[{{Maloney} {et~al.}(1996){Maloney}, {Hollenbach}, \&
  {Tielens}}]{Mal96}
{Maloney}, P.~R., {Hollenbach}, D.~J., \& {Tielens}, A.~G.~G.~M. 1996, \apj,
  466, 561, \dodoi{10.1086/177532}

\bibitem[{{Marconi} \& {Hunt}(2003)}]{Mar03}
{Marconi}, A., \& {Hunt}, L.~K. 2003, \apjl, 589, L21, \dodoi{10.1086/375804}

\bibitem[{{Marinucci} {et~al.}(2017){Marinucci}, {Bianchi}, {Fabbiano}, {Matt},
  {Risaliti}, {Nardini}, \& {Wang}}]{Mar17}
{Marinucci}, A., {Bianchi}, S., {Fabbiano}, G., {et~al.} 2017, \mnras, 470,
  4039, \dodoi{10.1093/mnras/stx1551}

\bibitem[{{Marinucci} {et~al.}(2013){Marinucci}, {Miniutti}, {Bianchi}, {Matt},
  \& {Risaliti}}]{Mar13}
{Marinucci}, A., {Miniutti}, G., {Bianchi}, S., {Matt}, G., \& {Risaliti}, G.
  2013, \mnras, 436, 2500, \dodoi{10.1093/mnras/stt1759}

\bibitem[{{Marinucci} {et~al.}(2012){Marinucci}, {Risaliti}, {Wang}, {Nardini},
  {Elvis}, {Fabbiano}, {Bianchi}, \& {Matt}}]{Mar12}
{Marinucci}, A., {Risaliti}, G., {Wang}, J., {et~al.} 2012, \mnras, 423, L6,
  \dodoi{10.1111/j.1745-3933.2012.01232.x}

\bibitem[{{Marinucci} {et~al.}(2015){Marinucci}, {Matt}, {Bianchi}, {Lu},
  {Arevalo}, {Balokovi{\'c}}, {Ballantyne}, {Bauer}, {Boggs}, {Christensen},
  {Craig}, {Gandhi}, {Hailey}, {Harrison}, {Puccetti}, {Rivers}, {Walton},
  {Stern}, \& {Zhang}}]{Mar15}
{Marinucci}, A., {Matt}, G., {Bianchi}, S., {et~al.} 2015, \mnras, 447, 160,
  \dodoi{10.1093/mnras/stu2439}

\bibitem[{{McClintock} {et~al.}(1979){McClintock}, {van Paradijs}, {Remillard},
  {Canizares}, {Koski}, \& {V{\'e}ron}}]{McC79}
{McClintock}, J.~E., {van Paradijs}, J., {Remillard}, R.~A., {et~al.} 1979,
  \apj, 233, 809, \dodoi{10.1086/157444}

\bibitem[{{McMullin} {et~al.}(2007){McMullin}, {Waters}, {Schiebel}, {Young},
  \& {Golap}}]{McM07}
{McMullin}, J.~P., {Waters}, B., {Schiebel}, D., {Young}, W., \& {Golap}, K.
  2007, Astronomical Society of the Pacific Conference Series, Vol. 376, {CASA
  Architecture and Applications}, ed. R.~A. {Shaw}, F.~{Hill}, \& D.~J. {Bell},
  127

\bibitem[{{Meijerink} \& {Spaans}(2005)}]{Mei05}
{Meijerink}, R., \& {Spaans}, M. 2005, \aap, 436, 397,
  \dodoi{10.1051/0004-6361:20042398}

\bibitem[{{Meijerink} {et~al.}(2007){Meijerink}, {Spaans}, \& {Israel}}]{Mei07}
{Meijerink}, R., {Spaans}, M., \& {Israel}, F.~P. 2007, \aap, 461, 793,
  \dodoi{10.1051/0004-6361:20066130}

\bibitem[{{Moran} {et~al.}(2007){Moran}, {Barth}, {Eracleous}, \&
  {Kay}}]{Mor07}
{Moran}, E.~C., {Barth}, A.~J., {Eracleous}, M., \& {Kay}, L.~E. 2007, \apjl,
  668, L31, \dodoi{10.1086/522697}

\bibitem[{{Mori} {et~al.}(2001){Mori}, {Tsunemi}, {Miyata}, {Baluta},
  {Burrows}, {Garmire}, \& {Chartas}}]{Mor01}
{Mori}, K., {Tsunemi}, H., {Miyata}, E., {et~al.} 2001, in Astronomical Society
  of the Pacific Conference Series, Vol. 251, New Century of X-ray Astronomy,
  ed. H.~{Inoue} \& H.~{Kunieda}, 576

\bibitem[{{Mulchaey} {et~al.}(1994){Mulchaey}, {Wilson}, {Bower}, {Heckman},
  {Krolik}, \& {Miley}}]{Mul94}
{Mulchaey}, J.~S., {Wilson}, A.~S., {Bower}, G.~A., {et~al.} 1994, \apj, 433,
  625, \dodoi{10.1086/174671}

\bibitem[{{Mundell} {et~al.}(2000){Mundell}, {Wilson}, {Ulvestad}, \&
  {Roy}}]{Mun00}
{Mundell}, C.~G., {Wilson}, A.~S., {Ulvestad}, J.~S., \& {Roy}, A.~L. 2000,
  \apj, 529, 816, \dodoi{10.1086/308318}

\bibitem[{{Mushotzky}(1982)}]{Mus82}
{Mushotzky}, R.~F. 1982, \apj, 256, 92, \dodoi{10.1086/159886}

\bibitem[{{Nagar} {et~al.}(1999){Nagar}, {Wilson}, {Mulchaey}, \&
  {Gallimore}}]{Nag99}
{Nagar}, N.~M., {Wilson}, A.~S., {Mulchaey}, J.~S., \& {Gallimore}, J.~F. 1999,
  \apjs, 120, 209, \dodoi{10.1086/313183}

\bibitem[{{Nelson} \& {Whittle}(1995)}]{Nel95}
{Nelson}, C.~H., \& {Whittle}, M. 1995, \apjs, 99, 67, \dodoi{10.1086/192179}

\bibitem[{{Nobukawa} {et~al.}(2010){Nobukawa}, {Koyama}, {Tsuru}, {Ryu}, \&
  {Tatischeff}}]{Nob10}
{Nobukawa}, M., {Koyama}, K., {Tsuru}, T.~G., {Ryu}, S.~G., \& {Tatischeff}, V.
  2010, \pasj, 62, 423, \dodoi{10.1093/pasj/62.2.423}

\bibitem[{{Pogge}(1989)}]{Pog89}
{Pogge}, R.~W. 1989, \apj, 345, 730, \dodoi{10.1086/167945}

\bibitem[{{Proga} {et~al.}(2014){Proga}, {Jiang}, {Davis}, {Stone}, \&
  {Smith}}]{Pro14}
{Proga}, D., {Jiang}, Y.-F., {Davis}, S.~W., {Stone}, J.~M., \& {Smith}, D.
  2014, \apj, 780, 51, \dodoi{10.1088/0004-637X/780/1/51}

\bibitem[{{Ramakrishnan} {et~al.}(2019){Ramakrishnan}, {Nagar}, {Finlez},
  {Storchi-Bergmann}, {Slater}, {Schnorr-M{\"u}ller}, {Riffel}, {Mundell}, \&
  {Robinson}}]{Ram19}
{Ramakrishnan}, V., {Nagar}, N.~M., {Finlez}, C., {et~al.} 2019, \mnras, 487,
  444, \dodoi{10.1093/mnras/stz1244}

\bibitem[{{Reynolds} \& {Mushotzky}(2017)}]{Ray17}
{Reynolds}, C.~S., \& {Mushotzky}, R. 2017, in AAS/High Energy Astrophysics
  Division \#16, AAS/High Energy Astrophysics Division, 103.23

\bibitem[{{Ricci} {et~al.}(2014{\natexlab{a}}){Ricci}, {Ueda}, {Ichikawa},
  {Paltani}, {Boissay}, {Gandhi}, {Stalevski}, \& {Awaki}}]{Ric14a}
{Ricci}, C., {Ueda}, Y., {Ichikawa}, K., {et~al.} 2014{\natexlab{a}}, \aap,
  567, A142, \dodoi{10.1051/0004-6361/201322701}

\bibitem[{{Ricci} {et~al.}(2015){Ricci}, {Ueda}, {Koss}, {Trakhtenbrot},
  {Bauer}, \& {Gandhi}}]{Ric15}
{Ricci}, C., {Ueda}, Y., {Koss}, M.~J., {et~al.} 2015, \apjl, 815, L13,
  \dodoi{10.1088/2041-8205/815/1/L13}

\bibitem[{{Ricci} {et~al.}(2014{\natexlab{b}}){Ricci}, {Ueda}, {Paltani},
  {Ichikawa}, {Gand hi}, \& {Awaki}}]{Ric14b}
{Ricci}, C., {Ueda}, Y., {Paltani}, S., {et~al.} 2014{\natexlab{b}}, \mnras,
  441, 3622, \dodoi{10.1093/mnras/stu735}

\bibitem[{{Ricci} {et~al.}(2017){Ricci}, {Trakhtenbrot}, {Koss}, {Ueda},
  {Schawinski}, {Oh}, {Lamperti}, {Mushotzky}, {Treister}, {Ho}, {Weigel},
  {Bauer}, {Paltani}, {Fabian}, {Xie}, \& {Gehrels}}]{Ric17nat}
{Ricci}, C., {Trakhtenbrot}, B., {Koss}, M.~J., {et~al.} 2017, \nat, 549, 488,
  \dodoi{10.1038/nature23906}

\bibitem[{{Rivers} {et~al.}(2014){Rivers}, {Markowitz}, {Rothschild}, {Bamba},
  {Fukazawa}, {Okajima}, {Reeves}, {Terashima}, \& {Ueda}}]{Riv14}
{Rivers}, E., {Markowitz}, A., {Rothschild}, R., {et~al.} 2014, \apj, 786, 126,
  \dodoi{10.1088/0004-637X/786/2/126}

\bibitem[{{Rosario} {et~al.}(2019){Rosario}, {Togi}, {Burtscher}, {Davies},
  {Shimizu}, \& {Lutz}}]{Ros19}
{Rosario}, D.~J., {Togi}, A., {Burtscher}, L., {et~al.} 2019, \apjl, 875, L8,
  \dodoi{10.3847/2041-8213/ab1262}

\bibitem[{{Rosario} {et~al.}(2010){Rosario}, {Whittle}, {Nelson}, \&
  {Wilson}}]{Ros10}
{Rosario}, D.~J., {Whittle}, M., {Nelson}, C.~H., \& {Wilson}, A.~S. 2010,
  \mnras, 408, 565, \dodoi{10.1111/j.1365-2966.2010.17153.x}

\bibitem[{{Shu} {et~al.}(2010){Shu}, {Yaqoob}, \& {Wang}}]{Shu10}
{Shu}, X.~W., {Yaqoob}, T., \& {Wang}, J.~X. 2010, \apjs, 187, 581,
  \dodoi{10.1088/0067-0049/187/2/581}

\bibitem[{{Shuder}(1980)}]{Shu80}
{Shuder}, J.~M. 1980, \apj, 240, 32, \dodoi{10.1086/158204}

\bibitem[{{Storchi-Bergmann} {et~al.}(1999){Storchi-Bergmann}, {Winge}, {Ward},
  \& {Wilson}}]{Sto99}
{Storchi-Bergmann}, T., {Winge}, C., {Ward}, M.~J., \& {Wilson}, A.~S. 1999,
  \mnras, 304, 35, \dodoi{10.1046/j.1365-8711.1999.02360.x}

\bibitem[{{Tanimoto} {et~al.}(2018){Tanimoto}, {Ueda}, {Kawamuro}, {Ricci},
  {Awaki}, \& {Terashima}}]{Tan18}
{Tanimoto}, A., {Ueda}, Y., {Kawamuro}, T., {et~al.} 2018, \apj, 853, 146,
  \dodoi{10.3847/1538-4357/aaa47c}

\bibitem[{{Togi} \& {Smith}(2016)}]{Tog16}
{Togi}, A., \& {Smith}, J.~D.~T. 2016, \apj, 830, 18,
  \dodoi{10.3847/0004-637X/830/1/18}

\bibitem[{{Tsunemi} {et~al.}(2001){Tsunemi}, {Mori}, {Miyata}, {Baluta},
  {Burrows}, {Garmire}, \& {Chartas}}]{Tsu01}
{Tsunemi}, H., {Mori}, K., {Miyata}, E., {et~al.} 2001, \apj, 554, 496,
  \dodoi{10.1086/321338}

\bibitem[{{Ulvestad} \& {Wilson}(1983)}]{Ulv83}
{Ulvestad}, J.~S., \& {Wilson}, A.~S. 1983, \apjl, 264, L7,
  \dodoi{10.1086/183935}

\bibitem[{{Urry} \& {Padovani}(1995)}]{Urr95}
{Urry}, C.~M., \& {Padovani}, P. 1995, \pasp, 107, 803, \dodoi{10.1086/133630}

\bibitem[{{Usero} {et~al.}(2004){Usero}, {Garc{\'\i}a-Burillo}, {Fuente},
  {Mart{\'\i}n-Pintado}, \& {Rodr{\'\i}guez-Fern{\'a}ndez}}]{Use04}
{Usero}, A., {Garc{\'\i}a-Burillo}, S., {Fuente}, A., {Mart{\'\i}n-Pintado},
  J., \& {Rodr{\'\i}guez-Fern{\'a}ndez}, N.~J. 2004, \aap, 419, 897,
  \dodoi{10.1051/0004-6361:20035774}

\bibitem[{{Wang} {et~al.}(2009){Wang}, {Fabbiano}, {Elvis}, {Risaliti},
  {Mazzarella}, {Howell}, \& {Lord}}]{Wan09}
{Wang}, J., {Fabbiano}, G., {Elvis}, M., {et~al.} 2009, \apj, 694, 718,
  \dodoi{10.1088/0004-637X/694/2/718}

\bibitem[{{Weaver} {et~al.}(1995){Weaver}, {Mushotzky}, {Serlemitsos},
  {Wilson}, {Elvis}, \& {Briel}}]{Wea95}
{Weaver}, K.~A., {Mushotzky}, R.~F., {Serlemitsos}, P.~J., {et~al.} 1995, \apj,
  442, 597, \dodoi{10.1086/175463}

\bibitem[{{Wilson} \& {Baldwin}(1985)}]{Wil85}
{Wilson}, A.~S., \& {Baldwin}, J.~A. 1985, \apj, 289, 124,
  \dodoi{10.1086/162870}

\bibitem[{{Young} {et~al.}(2001){Young}, {Wilson}, \& {Shopbell}}]{You01}
{Young}, A.~J., {Wilson}, A.~S., \& {Shopbell}, P.~L. 2001, \apj, 556, 6,
  \dodoi{10.1086/321561}

\end{thebibliography}

\end{document}